\documentclass[twocolumn,showpacs,aps,pre]{revtex4-2}

\usepackage{amssymb,amsfonts,amsmath}
\usepackage{euscript}
\usepackage{graphics}
\usepackage{setspace}
\usepackage{graphicx}
\usepackage{hyperref}%
\usepackage{setspace}
\usepackage{dcolumn}%
\usepackage{bm}%

\begin{document}

\title{Chemotactic aggregation dynamics of micro-swimmers in Brinkman flows}

\author{Yasser Almoteri$^{1}$, Enkeleida Lushi$^{2}$}
\email[]{lushi@njit.edu}
\affiliation{$^1$ Department of Mathematics and Statistics, Imam Mohammad Ibn Saud Islamic University, Riyadh 11623, Saudi Arabia\\
$^2$Mathematical Sciences, New Jersey Institute of Technology, Newark, NJ, 07102, United States}

\date{ April 28, 2025}

\begin{abstract}
We study through analysis and simulations of a continuum model the collective chemotactic dynamics of micro-swimmers immersed in viscous Brinkman flows. The Brinkman viscous flow approximates with a resistance or friction term the presence of inert impurities or stationary obstacles immersed in the fluid, an environment that can be regarded as a wet porous medium. Analysis of the linearized system reveals that resistance primarily affects the development of collective swimming instabilities and barely affects chemotactic instabilities. We present a parameter phase space for the distinct types of dynamics we can expect in the case of auto-chemotactic bacteria-like pusher swimmers for varying medium resistance, chemotactic response strength, and hydrodynamic coupling strength values. Simulations of the full nonlinear system show that resistance impacts the collective dynamics for each of these states because it inhibits the hydrodynamic interactions and the emergence of the collective swimmer. Surprisingly, and not expected from the linear analysis predictions, we find that resistance also hampers the chemotactic aggregation of the swimmers because it impedes their ability to navigate efficiently and collectively towards chemical cues and assemble into clusters. We show simulations of the complex system for parameters sets from each of the phase-space regions and quantify the observed behavior. Lastly, we discuss the experimental values of the parameters and discuss possible future experimental realizations of this system.
\end{abstract}

\maketitle


Microrganisms are one of the most abundant and diverse life forms, found in environments ranging from soils and ocean sediments to ice sheets and volcanic hot springs to even plant and animal tissues \cite{Pedley92, Lauga09, Ramaswamy10, Marchetti13, Bechinger16, Saintillan18, LiArdekani21, Spagnolie23}. Micro-swimmers such as bacteria and micro-algae are important in many natural processes ranging from fermentation, bioremediation and nitrogen fixation from the atmosphere, to technological process such as photo-bioreactors, drug production, or testing. Given their importance to life and their applications, studying the micro-swimmers’ behavior and their interactions with the environment is crucial to understanding the many phenomena they are involved in. Many experiments and theories have studied how such micro-swimmers move individually or collectively \cite{Dombrowski04, Kim04, Tuval05, Sokolov07, Sokolov09, Mino11, Cisneros11, Sokolov12}, in particular the varied dynamics that emerges when these swimmers ``communicate'' through fluidic disturbances or  chemical cues--the latter being the well-known chemotaxis process \cite{Berg72, Macnab72}. Macroscopic patterns are known to emerge as a result of these hydrodynamic or chemical interactions as well as collisions between the micro-swimmers, and significant work has been undertaken to understand the mechanisms that give rise to them, e.g. experiments, theories using continuum models, or direct particle simulations \cite{Simha02b, Hopkins02, Park03, Hill05, HernandezOrtiz05, Aranson07, Saintillan07, Saintillan08a, Ishikawa08, Baskaran09, Hernandez-Ortiz09, Subramanian09, Hohenegger10, Pedley10, Saragosti10, Saragosti11, Subramanian11, Saintillan11, Kasyap12, Lushi12, Ezhilan13, Dunkel13, Lushi13b, Lushi14, Kasyap14, Krishnamurthy15, Elgeti15, Lushi16, Wioland16, Stenhammar17, Lushi18, Skultety20, RojasPerez21, Murugan22, Traverso22, Fada22, Villa-Torrealba23}.

Though natural micro-swimmers live and have evolved in porous wet habitats such as soils, sediments, and tissues, not enough is known about their individual and collective motion in such environments. Most experiments have historically been conducted on homogenous environments such as bulk liquids or liquid layers above surfaces \cite{Bechinger16, LiArdekani21, Spagnolie23}. Experiments conducted in complex confinements such as large drops, racetracks, quasi-2D obstacle courses or 3D porous environments report that a non-trivial environment can significantly alter the individual and collective motion and transport of swimmers \cite{Volpe11,Majmudar12,Wioland13, Lushi14, Contino15, Sipos15, Wioland16, Nishiguchi18, Makarchuk19, Bhattacharjee19a, Kamdar22,Dehkharghani23}. For example, recent experiments have shown that tissue-inspired 3D pore-scale confinement not only affects micro-swimmer locomotion and migration but it also alters the chemotactic migration of the bacterial population \cite{Bhattacharjee19a, Bhattacharjee19, Bhattacharjee22}. Theoretical studies of bacterial chemotactic motion in wet porous environments have sometimes approximated the porosity effects on the colony's dynamics by modifying the motility parameters  in Keller-Segel chemotaxis models \cite{Ford07, Bhattacharjee21, Bhattacharjee22}. 

Theoretical and computational studies of individual and collective swimmer motion in complex environments and confinements have to confront the challenge of resolving  the hydrodynamical and collision interactions with other swimmers and any surfaces or obstacles or complexities present in their surroundings \cite{Wioland13, Lushi14, Spagnolie15, Datt15, Wioland16, Datt17, Desai17, Kos18,Nishiguchi18}. Continuum models, while more adept at handling population-level dynamics, cannot include the swimmer collisions, which are a crucial factor in determining their dynamics in complex confinement, though several approximations have been tried \cite{Ezhilan13, Bozorgi14, LiArdekani16, Desai17, Stoop19, LiArdekani21, Thijssen21, Kumar22}. Direct numerical simulations of the collective dynamics are often not feasible for large numbers of swimmers because of the difficulty and computational cost involved in properly resolving the non-local hydrodynamic interactions. Moreover, it is difficult to properly reconcile in direct simulations the directional force dipolar fluid flows that each swimmer generates by locomotion--a crucial component in the mechanism of collective swimming--and the quasi-instantaneous change in direction that results when micro-swimmers tumble--a crucial ability in processes such as chemotaxis. \\

A special case where the complex environment can be approximated is the Brinkman flow where a linear term is added to the momentum fluid equation to include the effect of stationary obstacles immersed in this fluid. Though seeming easy due to just the addition of the linear term, it is nonetheless non-trivial to determine the motion of single micro-swimmers through such a medium  \cite{Leshansky09, Sarah16, Ngangulia18, Nguyen19, Sarah20, Chen20}, and there had been no studies on the collective swimmer motion dynamics until our recent work \cite{Almoteri25a}. In said work, we presented a continuum model--constructed from swimmer micro-mechanics in Brinkman flow--that coupled the swimmer dynamics to that of the immersing Brinkman fluid. 

In the present work we extend that model to study the collective chemotactic dynamics of micro-swimmers in Brinkman flow, focusing on the case where the swimmers produce the chemo-attractant and swim towards it. The model consists of coupled equations for the swimmer configuration, the chemo-attractant field, and the Brinkman fluid equations with an extra active stress due swimmers moving in it. The swimmer conservation equation for the swimmer configuration distributions includes swimmer self-propulsion, which is modified by the Brinkman medium, advection and rotations by the local fluid flow, translational and rotational diffusions, as well as terms to model the run-and-tumble swimmer response to the changing chemo-attractant field. The chemo-attractant field, produced by the swimmers, is also dynamic and includes advection by the fluid and diffusion. Linear analysis of the uniform isotropic swimmer state about a quasi-static chemo-attractant field yields two separate dispersion relations, one stemming from chemotaxis and the other from the hydrodynamic collective swimming. The resistance appears in both due to the effect it has in slowing down individual swimmer motions; however its primary effect on the collective swimmer dynamics is through the fluid bulk-effect, namely, the linear frictional term in the fluid momentum equations. Moreover, the linear analysis helps us construct a phase diagram of the parameter space for the expected dynamics.  We perform numerical simulations of the full system for the dynamical states identified by the linear analysis, and they confirm that the medium's resistance or porosity hinders not only the collective swimming states but also the chemotactic aggregation behavior since it inhibits the swimmer motion and hydrodynamic interactions with others. Lastly, we present parameter values from the experiments that may be relevant to our system and discuss possible future experimental realizations.

\vspace{-0.3in}
\section{Swimming in Brinkman flow}
\vspace{-0.1in}

Micro-swimmers are naturally found in complex wet porous environments such as soils and tissue. In cases where such a fluidic environment contains  sparse stationary inert obstacles that are smaller in size than the flow's characteristic length scale \cite{Cortez10}, the fluid flow can be approximated using the {\it Brinkman Equations} \cite{Brinkman47, Brady87}, which incorporate a lower order resistance term into the viscous flow equations \cite{Nguyen19}:
\begin{align}
\mu \nabla^2 \mathbf{u} -\nabla q -(\mu / K_D)\mathbf{u}= 0, \quad \quad \quad
\nabla \cdot \mathbf{u}= 0
\end{align}
where $\mu$ is the viscosity, $\mathbf{u}$ is the fluid velocity, $q$ is the fluid pressure, and $K_D>0$ is the constant Darcy permeability  \cite{Cortez10, Vanni00}. The Green's function for the Brinkman equations is referred to as a Brinkmanlet  \cite{Cortez10} .

\begin{figure}[ht]
\centering
\includegraphics[width=3.4in]{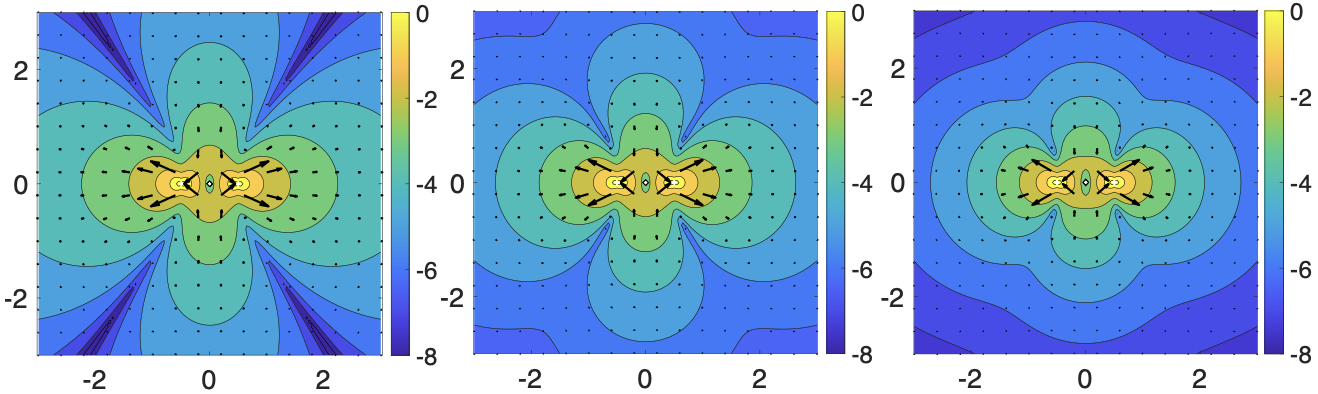}
\vspace{-0.3in}
\caption{Fluid velocity generated by two same-magnitude opposite-direction forces placed at distance $1$ apart in Brinkman flows with  resistances $\ell_p/K_D=0, 1, 2$, with $\ell_p$ the characteristic impurity size. The vector field indicates the fluid velocity $\mathbf{u}$, whereas the field color represents $\log |\mathbf{u}|$.}
\label{SwimmerFlow}
\end{figure}

In our forerunner work \cite{Almoteri25a} we showed how a rear-activated micro-swimmer, modeled as a bead-and-spring dumbbell  \cite{HernandezOrtiz05, Hernandez-Ortiz09}, moves through a Brinkman fluid medium with speed $U_B=U/(1+\nu+\nu^2/9)$, and moreover, the disturbance fluid flow that it generates due to its locomotion is a Brinkmanlet-dipole. Thus, not only does the medium porosity or friction presented by the embedded obstacles alter an individual swimmer's motion as seen in the speed correction, it also alters how this individual's motion affects other swimmers's motion through the hydrodynamic interactions because the swimmer's signature flow is altered, see Fig. \ref{SwimmerFlow}.

\vspace{-0.1in}
\vspace{-0.1in}

This figure illustrates the motivation of our study:  {\em (1) Given that the medium resistance alters the motion and flow disturbance generated by a swimmer \cite{Almoteri25a, Ngangulia18, Nguyen19, Sarah20}, and given that hydrodynamic interactions between swimmers are crucial in the emergence of macroscopic collective behavior in Stokes flow \cite{ Saintillan07, Saintillan08a, Saintillan08b, Lushi13b, Lushi14, Wioland16}, how is the collective swimmer motion altered when they are immersed in Brinkman flow? (2) Moreover, how does the Brinkman medium alter the collective chemotactic behavior of the micro-swimmers?}

The first question was studied in our forerunner work for non-tumbling micro-swimmers \cite{Almoteri25a}, and we will expand on it here for the case of tumbling swimmers. The second question is at the heart of this study.

\vspace{-0.3in}
\section{Mathematical Model}
\vspace{-0.1in}
\subsection{Continuum theory of micro-swimmers in Brinkman flow}
\vspace{-0.1in}

Following the model presented in our forerunner work \cite{Almoteri25a}, work on Stokesian swimmer suspensions  \cite{Saintillan08a, Saintillan08b, Subramanian09, Subramanian11, Lushi12, Lushi16, Lushi18}, and work on chemotactic response in bacterial populations \cite{Bearon00, Chen03, Subramanian11, Lushi12, Ezhilan12, Kasyap12, Kasyap14, Lushi16, Desai18}, we represent the configuration of ellipsoidal micro-swimmers by a distribution function $\Psi(\mathbf{x},\mathbf{p},t)$ of the center of mass position $\mathbf{x}$ and orientation vector $\mathbf{p}$ ($|\mathbf{p}|=1$). The suspension dynamics is described by the conservation equation:
\begin{align}
 \frac{\partial \Psi}{\partial t} = &- \nabla_x \cdot [ \Psi (U_B \mathbf{p} + \mathbf{u} ) ]  \nonumber \\
 &- \nabla_p \cdot [ \Psi  (I- \mathbf{p}\mathbf{p}^{ T})(\gamma \mathbf{E} + \mathbf{W}) \mathbf{p} ] \nonumber \\
 &+ \left[\Psi \lambda (\mathbf{p})  - \frac{1}{4\pi}\int \Psi  (\mathbf{p'}) \lambda (\mathbf{p'})  d \mathbf{p'} \right] \nonumber \\
  &+D \nabla_x^2 \Psi + d_r \nabla_p^2 \Psi. \label{ConEq}
\end{align}
The first term on the right hand side describes the swimmers' advection by the fluid with velocity $\mathbf{u}$ and self-propulsion with individual speed $U_B=U_0 h(\nu)$ along their orientation, with $U_0$ being the Stokesian swimmer speed and $h(\nu)=1/(1+\nu+\nu^2/9)$ being the correction incurred in a Brinkman medium \cite{Almoteri25a}. The second term describes swimmer rotation by the fluid, with $\gamma$ the aspect ratio ($0$ for spheres and $\approx 1$ for elongated rod-like swimmers), and $2\mathbf{E}= (\nabla \mathbf{u} + \nabla \mathbf{u} ^T )$, $2\mathbf{W}= (\nabla \mathbf{u} - \nabla \mathbf{u} ^T )$. The last two terms in Eq.  \ref{ConEq} describe translational and rotational diffusions with constant rates $D$ and $d_r$.

The third term in Eq. \ref{ConEq}, inside square brackets, describes run-and-tumble chemotaxis \cite{Berg72, Macnab72}. $\lambda (\mathbf{p})$ is the swimmer tumbling rate or frequency, describing the probability of the swimmer changing direction in a time interval, and it depends on the local concentration of the chemo-attractant field $C(\mathbf{x},t)$, as experienced by the swimmer. In this model we use a linearized biphasic form for the tumbling response \cite{Chen03, Lushi12}:
\begin{equation}
\mathbb{\lambda}(\mathbf{p})=
    \begin{cases}
        \lambda_0 \; (1 -\chi \mathcal{D}_t C) & \text{if } \mathcal{D}_t C > 1/\chi\\
         \lambda_0 & \text{otherwise}.
    \end{cases}
\end{equation}
\begin{align}
\mathcal{D}_t C= \frac{\partial C}{\partial t}+\mathbf{u} \cdot \nabla C+ U_B \mathbf{p} \cdot \nabla C \label{DtC}
\end{align}
$\mathcal{D}_t C$ is the chemo-attractant rate of change along the micro-swimmer path. $\lambda_0$ is the basic tumbling rate in the absence of chemotaxis. The chemotactic strength denoted by $\chi$ captures the swimmer's responsiveness to the chemo-attractant rate of change  $\mathcal{D}_t C$  \cite{Lushi12, Lushi16, Lushi18}. 

The chemo-attractant has its own dynamic behavior that encompasses advection by the fluid and diffusion with constant rate $D_c$:
\begin{align}\label{ChemEq}
 \frac{\partial C}{\partial t}+\mathbf{u} \cdot \nabla C = D_c \nabla^2 C - \beta_1 C + \beta_2 \Phi. 
\end{align}
The term -$\beta_1 C$ accounts for the degradation of chemo-attractant occurring at a constant rate $\beta_1$. The term $\beta_2 \Phi$ with constant rate $\beta_2$ captures the localized consumption ($\beta_2<0$) or production ($\beta_2>0$) of the chemo-attractant by the swimmers \cite{Lushi12, Lushi18}.  We focus on the {\em autochemotaxis} case where $\beta_2>0$ and the swimmers produce the chemoattractant.

The fluid velocity $\mathbf{u}(\mathbf{x},t)$ satisfies the Stokes-Brinkman equations with an extra active stress due to the swimmers moving in it:
\begin{align}\label{Stokes-Brin}
 -\mu \nabla_x^2 \mathbf{u} +\nabla_x q + \mu/K_D \mathbf{u} = \nabla_x \cdot \Sigma^p, \quad
  \nabla_x \cdot \mathbf{u} = 0.
\end{align}
 Here $\mu, q$ are the fluid viscosity and its pressure, $K_D$ the Darcy permeability constant for the porous medium \cite{Almoteri25a}, and $\Sigma^p$ is the active stress:
\begin{align}\label{stress-dim}
 \Sigma^p (\mathbf{x},t) = \sigma_0 \int \Psi (\mathbf{x},\mathbf{p},t) (\mathbf{pp}^T-\mathbf{I}/3)d\mathbf{p}.
\end{align}
The active stress $\Sigma^p$ is the average configuration of the active stresses $\sigma_0 (\mathbf{pp}^T-\mathbf{I}/3)$ exerted on the fluid by the swimmers over all possible orientations $\mathbf{p}$, with $\mathbf{I}/3$ added for symmetry \cite{Saintillan08a, Saintillan08b}. The dipole strength $\sigma_0$ is negative for a {\it pusher} swimmer with rear-activated propulsion like motile bacteria {\em E. coli} or spermatozoa, and it is positive for {\it puller} swimmers with front-activated propulsion like micro-algae {\em C. reinhardtii}  \cite{HernandezOrtiz05, Saintillan08a, Saintillan08b, Hernandez-Ortiz09}.

The local swimmer concentration $\Phi (\mathbf{x},t)$ and mean swimmer director director $<\mathbf{p}(\mathbf{x},t)>$ are defined as:
\begin{align}
 \Phi(\mathbf{x},t) &= \int \Psi (\mathbf{x},\mathbf{p},t) d \mathbf{p} , \label{PhiEqn} \\
 <\mathbf{p}(\mathbf{x},t)>  &=   \int d \mathbf{p} \mathbf{p} \Psi(\mathbf{x},\mathbf{p},t).
\end{align}

Note that the only changes of this continuum model for chemotactic swimmers in Brinkman flow from models for Stokes flow lie in the self-propulsion, namely the presence of the $U_B=U_0/(1+\nu + \nu^2/9)$ in the Eqs. \ref{ConEq}, \ref{DtC}, and most importantly through the presence of the resistance term in the fluid equations Eq. \ref{Stokes-Brin}.

\vspace{-0.1in}
\subsection{Non-dimensionalization of the system}\label{nondimsection}
\vspace{-0.1in}

We non-dimensionalize  the conservation Eq. (\ref{ConEq}) using the distribution, velocity, length and time scales $\Psi_c=n$, $u_c=U_0$, $l_c=(nl^2)^{-1}$, $t_c = l_c/u_c$, with $n$ being the mean number density of the micro-swimmers in a volume $V$. Here $V_p=Nl^3$ is the effective volume taken by $N$ swimmers of length $l$ in the fluid volume $V$ of a cube with length $L$ \cite{Saintillan08a}, so $l_c=(V/V_p)l$. This non-dimensionalization choice normalizes the distribution function as
$(1/V) \int_V d\mathbf{x} \int  \Psi(\mathbf{x},\mathbf{p},t) d\mathbf{p}= 1$, where $\Psi_0= 1/4\pi$ is the probability density for the uniform isotropic state.

Assuming a chemo-attractant scale $C_c$, the other non-dimensional parameters are $\lambda_0 = \overline{\lambda_0} t_c$, $\chi = \overline{\chi}C_c/t_c$, $\beta_1=\overline{\beta_1}t_c$, $\beta_2=\overline{\beta_2}n t_c/C_c$, $D_c=\overline{D_c} t_c/l_c^2$, $D=\overline{D} t_c/l_c^2$, $d_r = \overline{d_r}t_c$, with the overline denoting the dimensional parameters. 

We non-dimensionalize the fluid equations using $\alpha= \sigma_0 / ( U_0 \mu l_c^2 )$, and $\nu=  l_c/\sqrt{K_D}$. Here $\nu$ is the non-dimensional hydrodynamic resistance parameter, and $\alpha$ a non-dimensional $O(1)$ constant whose sign tells whether the micro-swimmers are bacteria-like pushers ($\alpha<0$) or micro-algae-like pullers ($\alpha>0$) \cite{Saintillan08a, Lushi18}. The non-dimensional Brinkman fluid equations with an active stress are:
\begin{align}\label{nondimfluid}
 &-\nabla_x^2 \mathbf{u} + \nabla_x q + \nu^2 \mathbf{u } = \nabla_x \cdot  \Sigma^p, \quad \quad \nabla_x \cdot \mathbf{u} = 0  \nonumber \\
 &\Sigma^p =\alpha  \int \Psi (\mathbf{x},\mathbf{p},t) [\mathbf{pp}^T-\mathbf{I}/3]d\mathbf{p}.
\end{align}

The non-dimensional chemo-attractant Eq.  \ref{ChemEq} remains unchanged and non-dimensional conservation Eq. \ref{ConEq} and chemo-attractant rate of change Eq. \ref{DtC} now have $U_B$ replaced by $h(\nu)$.

\vspace{-0.3in}
\section{Linear Stability}
\vspace{-0.1in}
\subsection{The eigenvalue problem} \label{TheEigenvalueProblem}
\vspace{-0.1in}

We study the stability of a swimmer suspension around the uniform and isotropic state $\Psi_0 = \frac{1}{4\pi}$. We assume that there is no swimmer rotational diffusion $(d_r = 0)$ and that the chemo-attractant field is quasi-static: 
\begin{align}\label{ChemoEqQuS}
D_c \nabla^2 C - \beta_1 C + \beta_2 \Phi =0. 
\end{align}

We focus on the case of auto-chemotactic bacteria-like pusher ($\alpha=-1$) swimmer suspensions when $\beta_1$ and $\beta_2$ are both positive, specifically when the swimmers produce the chemo-attractant. 

To analyze the dynamics of the system, we study the perturbations of the system around the uniform isotropic state $\Psi_0=1/4\pi$ for the swimmer distribution and steady-state for the chemo-attractant concentration $\bar{C}= \beta_2/\beta_1 \bar{\Phi} = \beta_2/\beta_1$.
\begin{align}
\Psi(\mathbf{x}, \mathbf{p},t) &=  \frac{1}{4\pi} + \epsilon {\Psi'}(\mathbf{x},\mathbf{p},t), C(\mathbf{x},t)=\frac{\beta_2}{\beta_1}+ \epsilon C'(\mathbf{x},t), \nonumber
\end{align}
with $|\epsilon| \ll 1$. With this choice, the tumbling rate is
\begin{align}
 \lambda(\mathcal{D}_t C)= \lambda_0 (1-\chi h(\nu) \mathbf{p} \cdot \nabla C). \nonumber  
\end{align}
The linearized distribution equation then is 
\begin{align}\label{LinDis}
\frac{\partial \Psi'}{\partial t}= &-\mathbf{p}^T \nabla \Psi'+ \frac{3\gamma}{4\pi}\mathbf{p^T} \mathbf{E}' \mathbf{p}  + D \nabla^2 \Psi' \nonumber \\
&-\lambda_0 \Psi'+\frac{\lambda_0 \chi h(\nu)}{4\pi} \mathbf{p}^T \nabla C'+ \frac{\lambda_0}{4\pi} \int \Psi' d \mathbf{p'}.
\end{align}
We consider plane-wave perturbations of the form $\Psi'(\mathbf{x}, \mathbf{p},t) = \tilde{\Psi}(\mathbf{p},\mathbf{k}) \exp (i \mathbf{k}^T\mathbf{x} + \sigma t)$, and similarly for $C'$ and $\mathbf{E}'$. Here $\mathbf{k}= k \hat{\mathbf{k}}$ is the wavenumber and $k= |\mathbf{k}|$.

We can solve Eq. (\ref{ChemoEqQuS}) for the chemo-attractant concentration $C$ and the fluid equations in terms of the swimmer distribution $\Psi$: 
\begin{align}
\tilde{C}&= \frac{\beta_2}{\beta_1+ k^2 D_c} \int \tilde {\Psi'} d \mathbf{p'}, \label{ConTe} \\
\tilde{\mathbf{u}} &= \frac{i k \alpha}{k^2 +\nu^2}(\mathbf{I} - \hat{\mathbf{k}}\hat{\mathbf{k}}^T) \left[  \int \tilde{\Psi'} \mathbf{p'p'}^T d\mathbf{p'} \right]  \hat{\mathbf{k}}.\label{teldaveloci} 
\end{align}

Substituting the expressions for $\nabla \tilde{\mathbf{u}}$ and $\tilde{C}$ in Eq. (\ref{LinDis}), we obtain:
\begin{align}
&\sigma  \tilde{\Psi}= -ikh(\nu) \mathbf{p}^T \hat{\mathbf{k}} \tilde{\Psi} -D k^2  \tilde{\Psi} \nonumber \\&- \frac{3 \gamma \alpha k^2 }{4\pi (k^2 + \nu^2)} \mathbf{p}^T (\mathbf{I} - \hat{\mathbf{k}}\hat{\mathbf{k}}^T) \int \mathbf{p'} \mathbf{p'}^T\tilde{\Psi} (\mathbf{p'}) d\mathbf{p'} \; \hat{\mathbf{k}}\hat{\mathbf{k}}^T  \mathbf{p} \nonumber \\
 &- \lambda_0 \tilde{\Psi}+\frac{\lambda_0}{4\pi} \; 
 \left(1+ \frac{\beta_2 \chi i k h(\nu) \mathbf{p}^T \hat{\mathbf{k}}}{\beta_1+ k^2 D_c}\right) \int \tilde{\Psi}(\mathbf{p'})d\mathbf{p'}.
\end{align}

Without loss of generality, we let $\mathbf{\hat{k}} = \mathbf{\hat{z}}$ and as $\mathbf{p}=\left[ \sin \theta \cos \phi, \sin \theta \sin \phi, \cos \theta \right]$
 for $\theta \in [0, \pi]$, $\phi \in [0, 2\pi)$, we can write: 
\begin{align}\label{EqSubsi}
&(\sigma+\lambda_0+ik h(\nu) \cos \theta) \tilde{\Psi} = \nonumber \\
&- \frac{3 \gamma \alpha k^2}{4\pi (k^2+\nu^2)} \cos \theta \sin \theta \left[ \cos \phi F_1(\Psi) + \sin \phi F_2(\Psi) \right] \nonumber \\
&+ \frac{\lambda_0}{4 \pi} \left[1+ \frac{\beta_2 \chi h(\nu)}{\beta_1+ k^2 D_c} \;  i k \cos \theta \right] G(\Psi), 
\end{align}
where for simplicity we defined the operators of $\tilde{\Psi}$:
\begin{align}
F_1(\tilde{\Psi})&= \int_0^{2\pi}  \cos \phi' \int_0^{\pi} \sin^2 \theta' \cos \theta' \tilde{\Psi}(\theta', \phi') d\theta' d\phi'  \nonumber \\
F_2 (\tilde{\Psi})&=\int_0^{2\pi}  \sin \phi' \int_0^{\pi} \sin^2 \theta' \cos \theta' \tilde{\Psi}(\theta', \phi') d\theta' d\phi'  \nonumber \\
G(\tilde{\Psi})&=\int_0^{2\pi} \int_0^{\pi} \sin \theta' \tilde{\Psi}(\theta', \phi') d\theta' d\phi'. \nonumber
\end{align}

To solve for the growth rate $\sigma$, we apply the operators $F_1$ and $G$ to $\tilde{\Psi}$ in Eq.(\ref{EqSubsi}) and this leads to two uncoupled integral dispersion relations:
\begin{align}
1&= - \frac{3 \alpha k^2}{4\pi (k^2+\nu^2)} \int_0^{\pi} \frac{\sin^3 \theta \cos^2 \theta d\theta}{(\sigma+\lambda_0+ik h(\nu) \cos \theta + Dk^2)}  \nonumber \\
1&= \frac{\lambda_0}{2}  \frac{\chi \beta_2 h(\nu) i k }{(\beta_1+ k^2 D_c)}  \int_0^{\pi} \frac{\sin \theta \cos \theta d\theta}{(\sigma+\lambda_0+ik h(\nu) \cos \theta + Dk^2)}  \nonumber \\
&+\frac{\lambda_0}{2} \int_0^{\pi} \frac{\sin \theta  d\theta}{(\sigma+\lambda_0+ik h(\nu) \cos \theta + Dk^2)}.  \nonumber
\end{align}

Letting $a=(\sigma  + \lambda_0 + Dk^2)/(ik h(\nu))$, $R= \chi \beta_2/(\beta_1+ k^2 D_c)$ and evaluating the integrals leads to two  dispersion relations that are implicit equations for $\sigma(k)$:
\begin{align}
1 &= -\frac{3 \gamma \alpha k^2}{4(k^2 + \nu^2)} \frac{1}{ik} \left[  2a^3 -\frac{4}{3}a +(a^4-a^2)\log \left( \frac{a-1}{a+1} \right)  \right],  \label{Hrelation2}\\
1 &=  \frac{\lambda_0}{2} R h(\nu) \left[2+ a \log \left(\frac{a-1}{a+1} \right) \right] - \frac{\lambda_0}{2} \frac{1}{ik} \log \left(\frac{a-1}{a+1} \right). \label{RTrelation}
\end{align} 

We will refer to Eq. (\ref{Hrelation2}) as the {\it hydrodynamic dispersion relation} due to the presence of hydrodynamics-related parameters $\gamma, \alpha, \nu$, and we refer to Eq. (\ref{RTrelation}) as the {\it chemotaxis dispersion relation} based on the presence of chemotaxis-related parameters $\chi, \beta_1, \beta_2$. 

The hydrodynamic dispersion relation has the same form as that found for non-chemotactic swimmers in Brinkman flow, explained in our forerunner work \cite{Almoteri25a}, except now the term $a$ includes the tumbling rate $\lambda_0$. 

The auto-chemotactic dispersion Eq. \ref{RTrelation} is similar in form to that found by Lushi et al. \cite{Lushi12, Lushi16, Lushi18} for auto-chemotactic micro-swimmer suspensions in Stokes flow, but here in Brinkman flow we see the appearance of the correction $h(\nu)$. This relation is unaffected by the swimming mechanism or swimmer geometry as noted by the absence of the dipole strength  $\alpha$ and the swimmer shape $\gamma$. The auto-chemotactic relation relies primarily on the chemo-attractant dynamics, as indicated by the term $R$.

\vspace{-0.2in}
\subsection{Asymptotic solutions of the dispersion relations}
\vspace{-0.1in}

The dispersion relations described by Eqs. \ref{Hrelation2}, \ref{RTrelation} cannot be solved analytically for the growth rate $\sigma(k)$, so we seek small $k$ asymptotic solutions $\sigma(k)=\lambda_0 + \sigma_0+\sigma_1 k+\sigma_2 k^2+O(k^3)$ for both dispersion relations \cite{Almoteri25a}.

For the hydrodynamic dispersion relation Eq. \ref{Hrelation2}, we obtain two branches for the growth rate:
\begin{align}
\sigma_{H_1} &= -\lambda_0+ \frac{ (-\alpha \gamma)k^2}{5(k^2+\nu^2)}h(\nu) - [\frac{15(k^2+\nu^2)}{7(-\alpha \gamma)}+D]k^2 +...  \label{Hrelationexpan.1}\\
\sigma_{H_2} &= -\lambda_0+ \frac{(k^2+\nu^2)}{(-\alpha \gamma)} h(\nu)-D k^2 +...  \label{Hrelationexpan.3}
\end{align}

The two solutions in Eqs. \ref{Hrelationexpan.1}, \ref{Hrelationexpan.3} are similar to those found  for non-tumbling non-chemotactic swimmers in Brinkman flow \cite{Almoteri25a}, but now both branches decrease by the basal tumbling $\lambda_0$. Tumbling per se has a stabilizing effect on the suspension, as does translational diffusion. 

Both branches of $\sigma_{H}(k)$ are negative for puller swimmers ($\alpha>0$) for any basal tumbling value and any swimmer shape, which means the puller suspension perturbation is stable. For pushers ($\alpha<0$), it may be possible for $\sigma_{H_1}$ to be positive for some parameters and wave-numbers, especially if it can overcome $\lambda_0$--e.g. see plots of Eqs. \ref{Hrelationexpan.1}, \ref{Hrelationexpan.3} for non-tumbling swimmers in  \cite{Almoteri25a}.  

In \cite{Almoteri25a} we showed that for non-tumbling swimmers, $\lambda_0=0$, resistance has a stabilizing effect as both $\sigma_{H1}(k)$ and $\sigma_{H2}(k)$ decrease with increasing $\nu$. Here that is also the case since both branches just decrease by $\lambda_0$. Note that for Stokesian swimmers $\sigma_{H_1}(k=0, \nu=0) = (-\alpha \gamma)/5 - \lambda_0$, and for Brinkman swimmers $\sigma_{H_1}(k=0, \nu \neq 0) = 0 - \lambda_0$, thus there's a discontinuity at $k=0$. 

For the autochemotactic relation Eq. (\ref{RTrelation}), as in the Stokes case \cite{Lushi12, Lushi16, Lushi18}, we obtain only one branch for the growth rate:
\begin{align}\label{RTrelationexapn}
&\sigma_{C} = \lambda_0(h(\nu)-1) + h(\nu)\frac{ \lambda_0  \bar{\chi} h(\nu) -1}{3\lambda_0} k^2 \nonumber \\
&- [\frac{ \bar{\chi} \bar{D_c}h^2(\nu)}{3} + h(\nu)\frac{(5\lambda_0  \bar{\chi}h(\nu)  -1)(   \lambda_0 \bar{\chi} h(\nu)  -1)}{45 \lambda^3_0}] k^4 +...\\
& \text{where } \bar{\chi} :=(\chi \beta_2)/\beta_1, \quad \bar{D_c}:=D_c / \beta_1.\nonumber
\end{align}

Eq. \ref{RTrelationexapn} shows that there's a possibility for $\sigma_{C}>0$ for some $k$ when $\lambda_0  \bar{\chi} h(\nu) -1>0$, or $\chi \beta_2/\beta_1 > 1/(\lambda_0 h(\nu))$.

\vspace{-0.2in}
\subsection{Numerical solutions of the dispersion relations}
\vspace{-0.1in}

We can solve the dispersion relations Eqs. (\ref{Hrelation2}) and (\ref{RTrelation}) numerically using with an iterative quasi-Newton solver with asymptotic results Eqs. \ref{Hrelationexpan.1}, \ref{Hrelationexpan.3}  as initial guesses --the procedure is explained in \cite{Almoteri25a, Almoteri23}. We focus on the case of {\em pusher} swimmers ($\alpha=-1$).

\begin{figure}[htbp]
\centering
\includegraphics[width=2.9in]{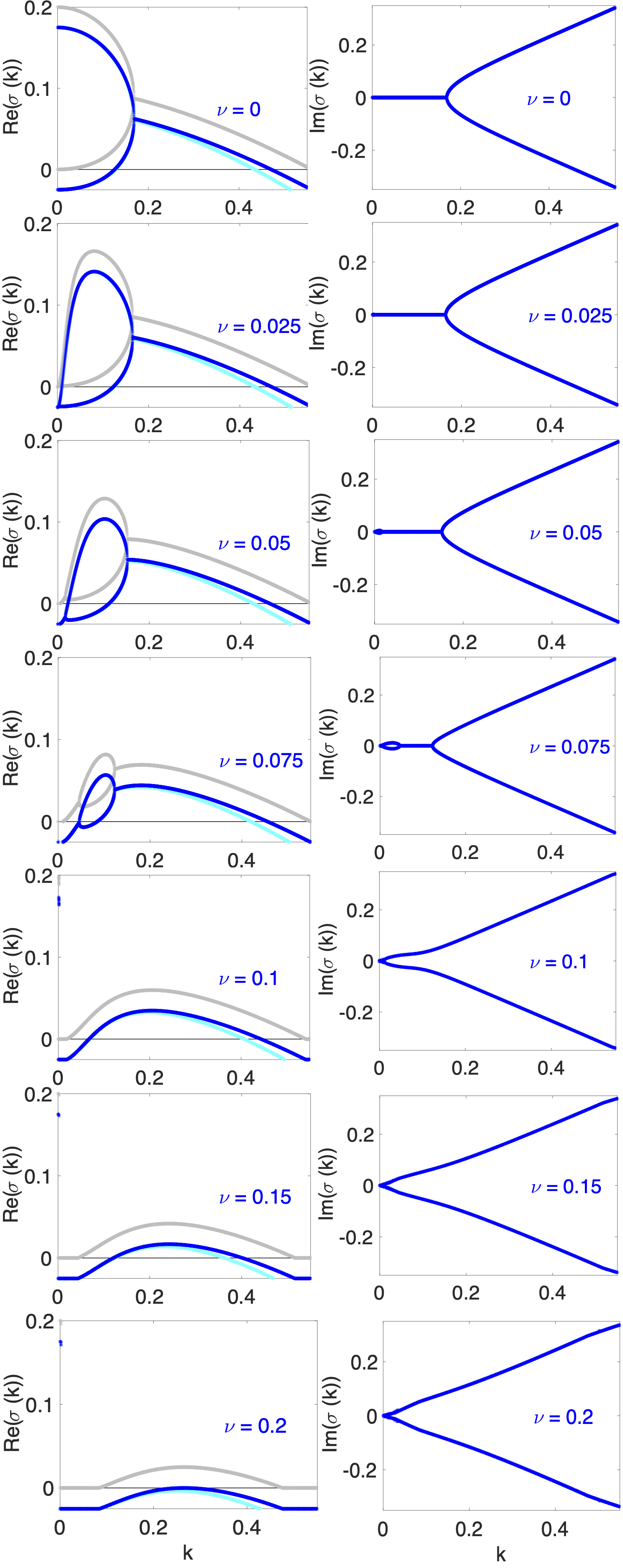}
\vspace{-0.1in}
\caption{Numerical solution of the dispersion relation Eq. (\ref{Hrelationexpan.1}) with $\lambda_0=0$ (in gray), $\lambda_0=0.025$ (in blue for $D=0$, in cyan for $D=0.05$) for various resistance values $\nu=0, 0.05,..., 0.25$ in the case of elongated pusher swimmers ($\alpha=-1, \gamma=1$). 
}
\label{fig:ReImSigma_RT}
\end{figure}

In \cite{Almoteri25a}, we discuss at length the rich behavior of the growth rate functions for the case of non-tumbling swimmers, which we will briefly summarize here. There are two branches of the growth rate, as predicted by the asymptotic analysis for small $k$ in Eqs. \ref{Hrelationexpan.1} \ref{Hrelationexpan.3}, and for larger $k$ the branches can develop  imaginary parts. Just as in the asymptotic results, $\sigma_{H1}(k=0,\nu=0)=1/5$ whereas $\sigma_{H1}(k=0,\nu \neq 0)=0$. At the resistance value $\nu_m\approx 0.085$, the two branches merge. 

There is a range of wave-numbers for which $\sigma_H>0$, and it is maximal for Stokesian swimmers $k \in (0, 0.57)$; this range gets smaller for increasing resistance values $\nu$ until it disappears altogether at the critical $\nu_c \approx 0.279$. 

For the same wavenumber $k$, it can be seen that $\nu_1<\nu_2$ gives $Re(\sigma_H(k, \nu_1))< Re(\sigma_H(k, \nu_1))$, thus overall the resistance has a stabilizing effect on the suspension. 

{\em Note that the linear theory says that  $\nu>\nu_z$ will turn off the hydrodynamic instability for pusher suspensions for {\it any} wave-numbers $k$, thus any system size $L=2\pi/k$.}

Tumbling affects the hydrodynamic instability in the Brinkman medium in a similar fashion to how it affects it in a Stokes medium: $Im(\sigma_H(k)$ is unaffected, but both branches of $Re(\sigma_H(k))$ are shifted down by $\lambda_0$ in comparison to the non-tumbling-cases. This shows that tumbling per se has a stabilizing effect on the system. 

{\em Note that $\lambda_0 \geq 0.2$ shifts the entire $Re(\sigma_H(k))$ to non-positive values, thus $\lambda_0 \geq 0.2$ suffices to turn off the hydrodynamic instability for {\it any} wavenumber $k$.}

From Eqs. \ref{Hrelationexpan.1} \ref{Hrelationexpan.3} and Fig. \ref{fig:ReImSigma_RT}, translational diffusion with constant rate $D$ is stabilizing for any type of suspension. Though not included here, we expect a stabilizing effect from rotational diffusion \cite{Hohenegger10}.

\begin{figure}[htbp]
\centering
\includegraphics[width=3.5in]{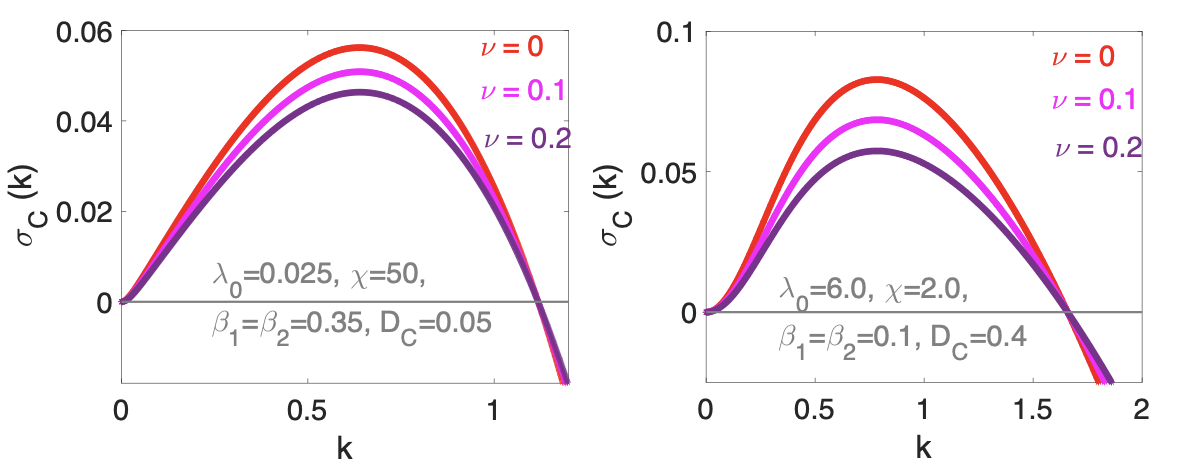}
\vspace{-0.3in}
\caption[Numerical solution of the auto-chemotactic dispersion relation for two sets of parameters.]{Numerical solution for the growth rate $\sigma_C (k)$ of the auto-chemotactic dispersion relation Eq. (\ref{RTrelation}) for two sets of parameters $\chi, \beta_1, \beta_2,D_c$. 
} 
\label{fig:Sigma_C}
\end{figure}

Fig. \ref{fig:Sigma_C} shows the numerical solution for $\sigma_C (k)$ of Eq. (\ref{RTrelation}) for two different parameter sets $\lambda_0,\chi, \beta_1, \beta_2,D_c$.  We notice $\sigma_C(k)$ is real-valued, $\sigma_C(0)=0$, and, consistently with the asymptotic results of Eq. (\ref{RTrelationexapn}), $\sigma_C(k)>0$ for small $k$ if $\chi \beta_2/\beta_1- 1/\lambda_0h(\nu)>0$. The initially-upward curve of $\sigma_C (k)$ turns and comes down to zero and eventually becomes negative for some larger $k$-value because chemo-attractant diffusion has a stabilizing effect at $O(k^4)$. In all, for auto-chemotactic swimmer suspensions, there is a finite band of unstable modes when $\chi \beta_2/\beta_1> 1/\lambda_0 h(\nu)$, and its width is controlled by chemo-attractant diffusion, as in Stokesian suspensions \cite{Lushi12}. 

The presence of hydrodynamic resistance here just serves to stabilize the system, as seen in Fig. \ref{fig:Sigma_C} because given any wavenumber $k$, we have $\sigma_C(k,\nu_2) < \sigma_C(k,\nu_1) < \sigma_C(k,\nu=0)$  for $\nu_2>\nu_1>0$.

These linear stability results say that the chemotactic growth rate is unaffected by the swimmer type and the collective motion emerging due to the hydrodynamic interactions; however, as we'll discover with simulations of the full nonlinear system, this is only true when the system is close to the uniform isotropic state.

\begin{figure}[htbp]
\centering
\includegraphics[width=2.1in]{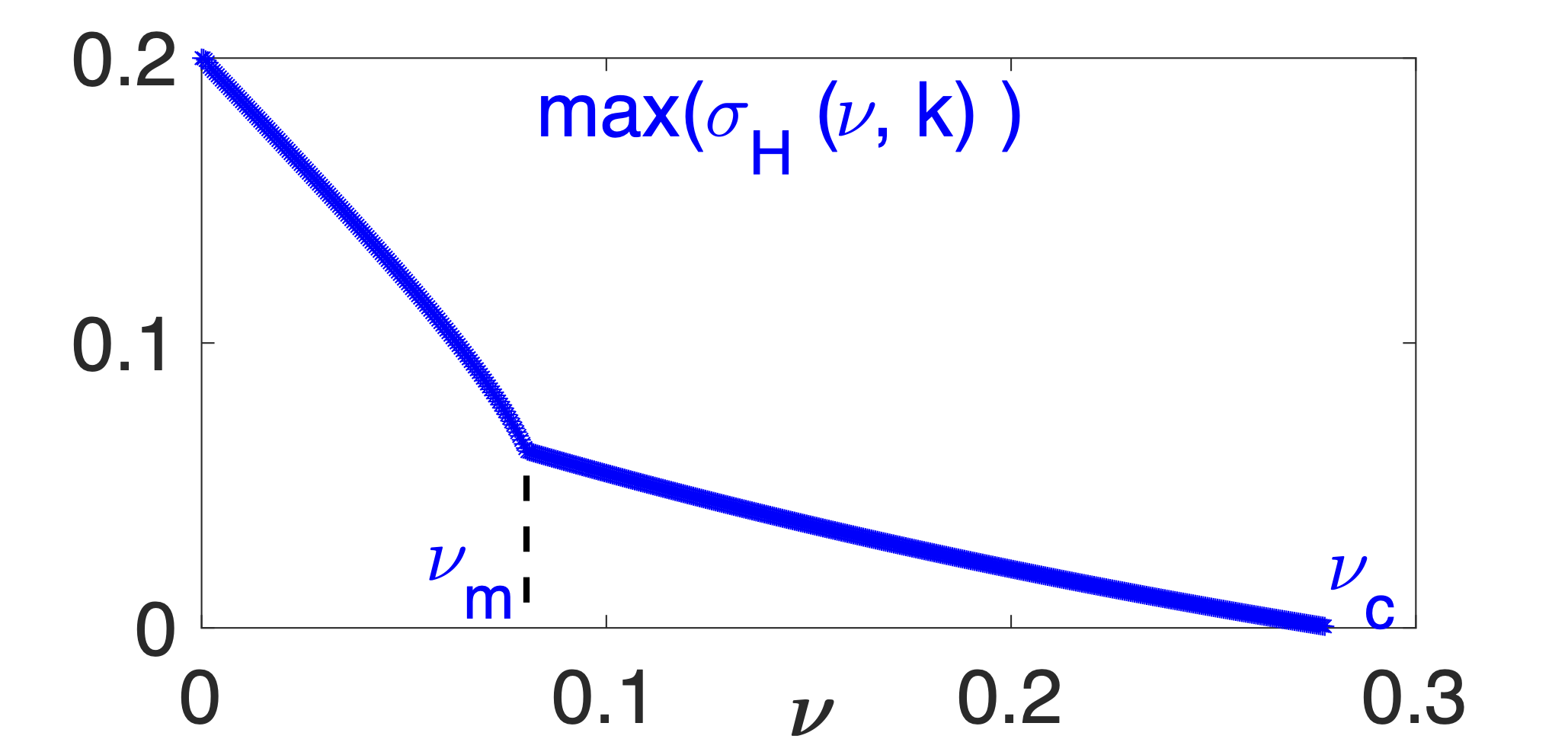}
\vspace{-0.1in}
\caption{ Maximum of $\Re(\sigma_H(k))$ for various parameters $\nu$. 
} 
\label{fig:maxsigma}
\end{figure}

Lastly, we plot in Fig.\ref{fig:maxsigma} for various resistance parameters the maximum of $Re(\sigma_{H1}(k))$, the real part of the hydrodynamic instability growth rate branch. 

\vspace{-0.1in}
\subsection{Phase diagram of complex dynamics}
\vspace{-0.1in}

The asymptotic analysis and the numerical solutions to the dispersion relations provide valuable information about parameter ranges in pusher suspensions:
\begin{itemize}
\item From Eq. (\ref{Hrelationexpan.1}) and Fig. \ref{fig:ReImSigma_RT} we see that $\lambda_0 \geq 0.2 = \max(-\alpha \gamma/5)$ turns off the hydrodynamic instability for any wavenumber, hence system size \cite{Lushi12}. 
\item From Eq. (\ref{RTrelationexapn}), to have a chemotactic instability $\sigma_{C}(k)>0$, we need  $\chi \beta_2/\beta_1> 1/\lambda_0 h(\nu)$. 
\item From Fig. \ref{fig:ReImSigma_RT} and Fig. \ref{fig:maxsigma}, and Eq. (\ref{RTrelation}), we see that picking $\lambda_0 \geq \lambda_0(\nu):= \max(Re(\sigma(k,\nu)))$ from the plot effectively disables the hydrodynamic instability in pushers for any system size. 
\end{itemize}

\begin{figure}[htpb]
\centering
\vspace{-0.1in}
\includegraphics[width=3.3in]{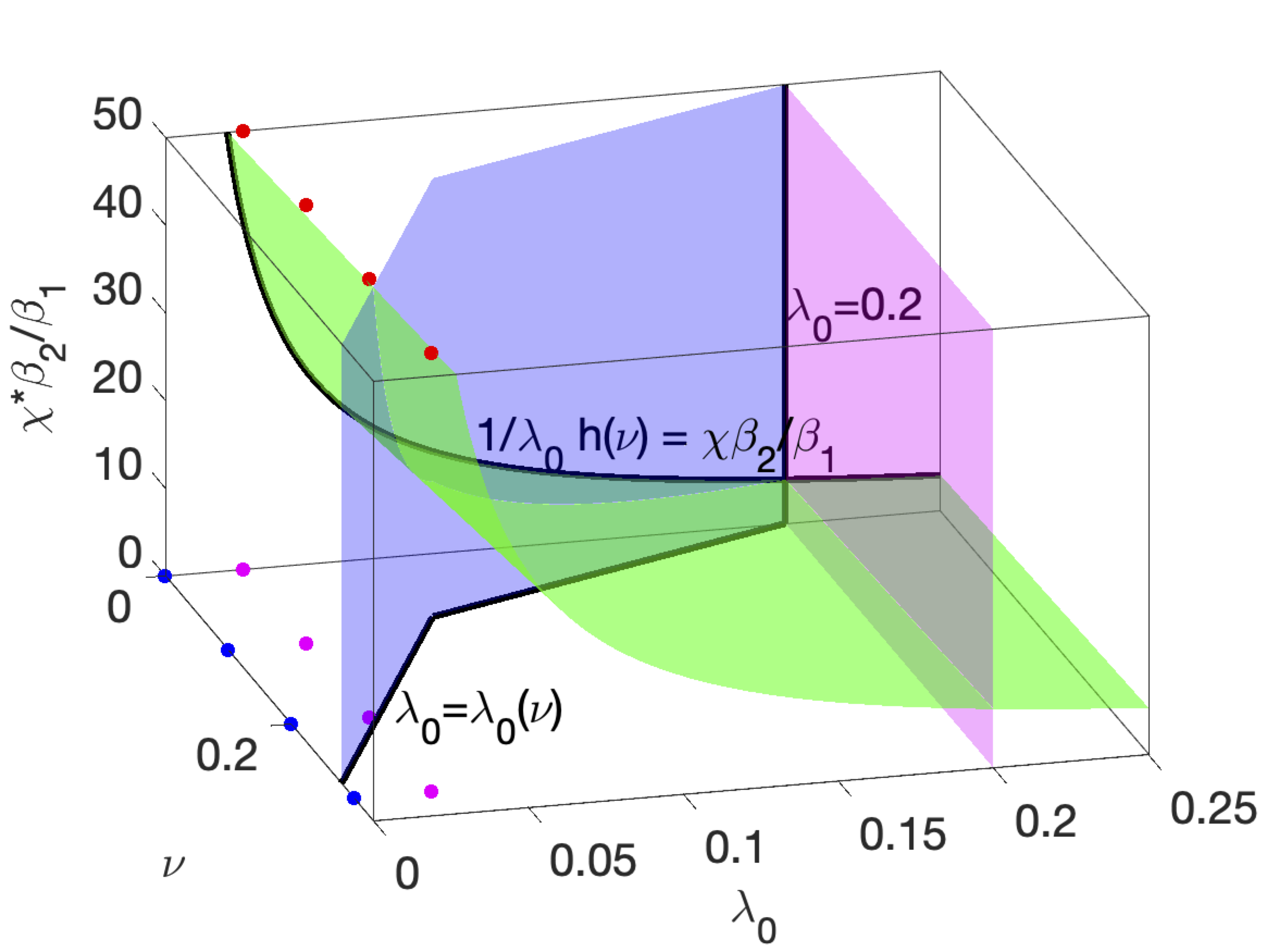}
\vspace{-0.2in}
\caption{3D phase diagram of parameter regions for expected dynamics depending on the presence of resistance, hydrodynamic and auto-chemotactic instabilities. The dots note parameter values utilized in nonlinear simulations.
}
\label{fig:PhaseDiagram_3D}
\end{figure}

Building upon this information gathered from the linear theory, we plot the surfaces $\lambda_0=0.2$,   $\chi \beta_2/\beta_1 = 1/\lambda_0  h(\nu)$ and $\lambda_0=\lambda_0(\nu)$ on a 3D plot with axis $\nu$, $\lambda_0$, and $\chi \beta_2/\beta_1$, shown in Fig. \ref{fig:PhaseDiagram_3D}. By examining the 3D regions resulting from the surfaces' intersections and where a parameter set is situated with respect to them, we can gain insight on which instabilities would likely arise, and thus gain insight on the combined effects of the hydrodynamics, auto-chemotaxis and resistance. 

\begin{figure}[htpb]
\centering
\includegraphics[width=2.8in]{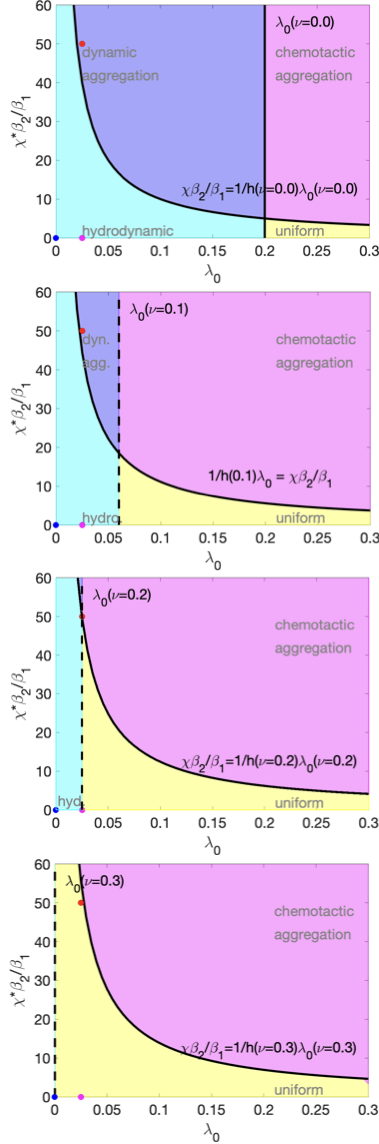}
\vspace{-0.2in}
\caption{Slices of the 3D phase diagram for $\nu=0, 0.1, 0.2, 0.3$ showing the change in the parameter regions for the four dynamical states. The dots note parameter values utilized in nonlinear simulations.
}
\label{fig:Phase_slices1}
\end{figure}

To better visualize these 3D regions, we show 2D slices of the phase diagram for various resistance parameters $\nu=0, 0.1, 0.2, 0.3$  in Fig. \ref{fig:Phase_slices1}. These slices help us comprehend how the system's dynamics evolve under different resistance conditions.

The phase diagram obtained for the case where $\nu=0.0$ and shown in Fig. \ref{fig:Phase_slices1}a is that found by Lushi  et al., \cite{Lushi12, Lushi16, Lushi18} for auto-chemotactic pusher suspensions in homogenous viscous fluids. This diagram reveals four distinct regions of dynamics, which were labeled as: 
\begin{itemize}
\item hydrodynamic collective swimming ($\lambda_0<0.2$, $\chi \beta_2/\beta_1<1/\lambda_0 h(\nu)$) with only the hydrodynamic instability, 
\item chemotactic aggregation ($\lambda_0>0.2$, $\chi\beta_2/\beta_1>1/\lambda_0 h(\nu)$) with only the chemotactic instability, 
\item dynamic aggregation or mixed dynamics ($\lambda_0<0.2$, $\chi \beta_2/\beta_1>1/\lambda_0 h(\nu)$) with both instabilities, 
\item uniform state ($\lambda_0>0.2 $, $\chi\beta_2/\beta_1<1/\lambda_0 h(\nu)$) with neither instability.
\end{itemize}

When resistance $\nu>0$ is introduced, a hydrodynamic instability can still occur if $\lambda_0 < \lambda_0(\nu)$, where $\lambda_0 (\nu)$ is defined as the maximum value of $\Re(\sigma_H(\nu))$ obtained from the curve in Fig. \ref{fig:maxsigma}. Essentially, as $\nu$ increases, the boundary for the hydrodynamic instability, which was initially at  $\lambda_0 (\nu=0)= 0.2$, shifts to the left. This can be observed in the Fig. \ref{fig:Phase_slices1}, where the regions corresponding to hydrodynamic collective swimming and dynamic aggregation shrink with increasing $\nu$. Eventually, these regions disappear entirely for critical resistance $\nu_c \approx 0.279$, as indicated by $\max(\sigma_H \nu \approx 0.279) = 0$. 

The stability criterion used to construct the phase diagram was $\lambda_0 (\nu)= \max(Re(\sigma(\nu)))$, which is valid for all $k$. A more accurate criterion can be derived if the specific domain or length-scale size $L$ is known apriori, $\lambda_0 (\nu) = Re(\sigma (\nu=0.2, k=2\pi/L))$.

Finally, the dots in Figs. \ref{fig:PhaseDiagram_3D} and \ref{fig:Phase_slices1} denote the parameter values employed in the nonlinear simulations. The blue dots represent the parameters for non-tumbling non-chemotactic swimmers that were studied in \cite{Almoteri25a}, while the magenta and red dots correspond to the simulations to be shown in the next section.

\vspace{-0.1in}
\section{Simulations of the Nonlinear System}
\vspace{-0.1in}

We investigate through numerical simulations of the full nonlinear system the impact of resistance on the dynamics regions identified through the linear analysis and depicted in Fig. \ref{fig:Phase_slices1}, namely the hydrodynamic collective swimming, the dynamic aggregation, and the auto-chemotactic aggregation states. 

We use the numerical method of our prior studies \cite{Almoteri25a}, briefly summarized here. For feasibility, we look at a periodic 2D system where swimmers are constrained in the $(x,y)$-plane and have one orientation angle $\theta \in [0, 2\pi)$, so $\mathbf{p}= (\cos\theta, \sin\theta,0)$. The domain is discretized uniformly with $M=128$-$256$ points in the $x$ and $y$ directions, and $M_{\theta}=16$-$32$ points in the angle/orientation. 

Second order accurate finite differences are used to calculate the flux terms in the $\Psi$ conservation equation, and the trapezoidal summation is used for the integrals in  $\theta$, e.g. for $\Sigma^p$ and $\Phi$.  The computational domain is periodic, so we employ spectral methods (Fast Fourier Transforms), to solve the fluid equations similar to the formula in Eq. (\ref{teldaveloci}). A second order Adams-Bashforth scheme is used to integrate Eq. \ref{ConEq} in time, with sufficiently small time-step to keep the calculations stable. 

The initial condition of the swimmers is chosen to be a random perturbation around the uniform isotropic state, as in forerunner studies \cite{Saintillan08a, Lushi12, Lushi18, Almoteri25a}. The initial chemo-attractant distribution is taken to be $C(\mathbf{x},0)= \beta_2 / \beta_1$. These initial conditions correspond to the case we studied through  linear analysis in the previous Section.

For this study, we choose a periodic square box with side $L = 25$ which allows for enough unstable modes to study the effect of resistance. Specifically, $k=2\pi/L \approx 0.25$ has $Re(\sigma_H(k))>0$ for a wide range of resistance values $\nu < \nu_c=0.279$ according to our linear analysis plots in Fig. \ref{fig:ReImSigma_RT}. We will examine other simulation box sizes in subsequent studies. 

Lastly, we focus here on {\em elongated pusher swimmers} $\gamma=1$, $\alpha=-1$. Translational and rotational diffusions are included with coefficients $D = d_r = 0.01$.

\vspace{-0.1in}
\subsection{Resisting hydrodynamic collective swimming}
\vspace{-0.1in}

To further explore the influence of resistance on the hydrodynamic collective swimming state, we consider tumbling but non-chemotactic pusher suspensions with basic tumbling rate $\lambda_0=0.025$. The parameter set is noted as magenta dots in the phase diagrams of Figs. \ref{fig:PhaseDiagram_3D} and \ref{fig:Phase_slices1}. From Fig. \ref{fig:ReImSigma_RT}, a hydrodynamic instability for this tumbling rate is expected for $\nu=0, 0.1$ but not for $\nu=0.2$. 

\begin{figure*}[ht]
\centering
\includegraphics[width=1.8\columnwidth]{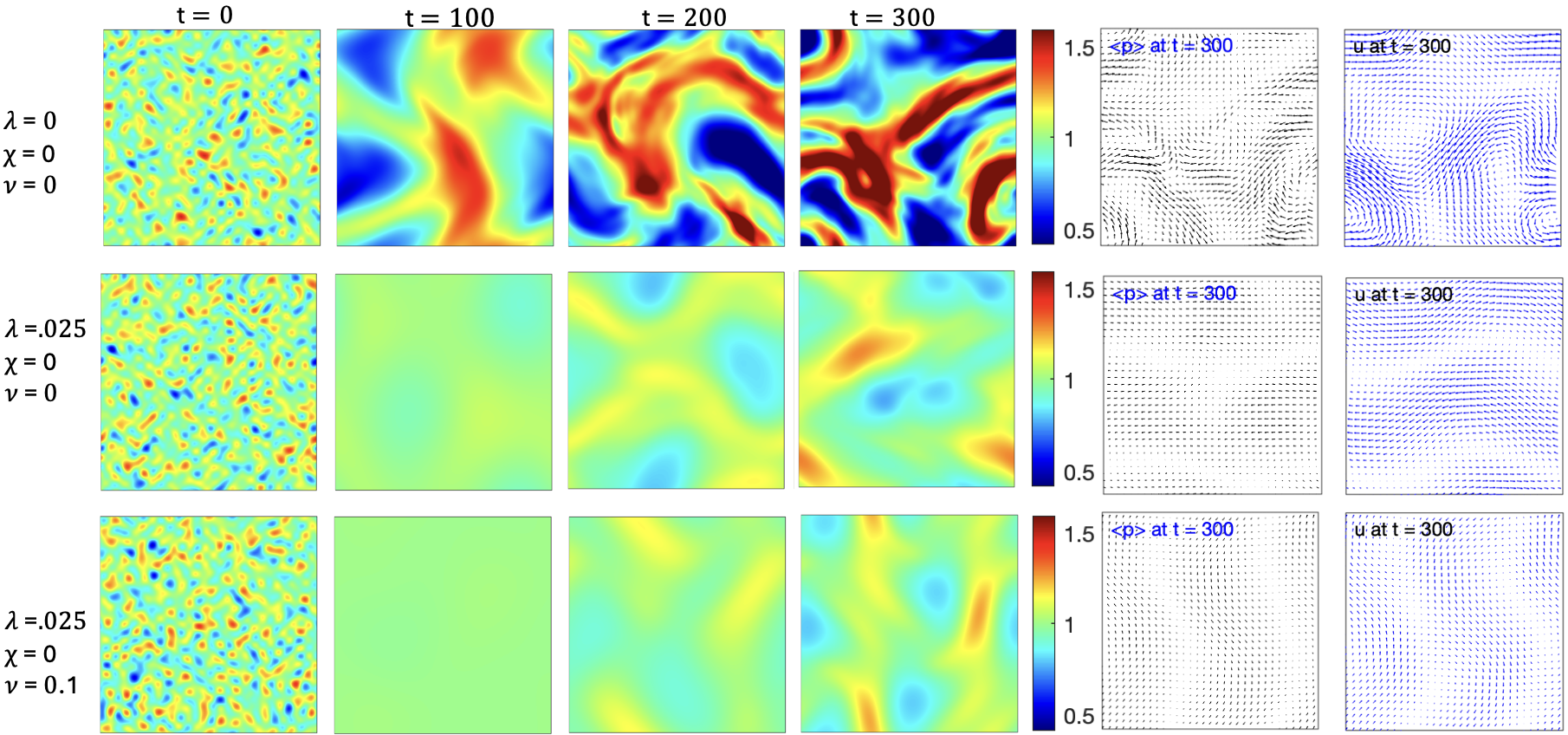}
\vspace{-0.1in}
\caption{The {\bf pusher} swimmer concentration $\Phi$ at times $t=100,200,300$, the swimmer director $<\mathbf{p}>$, the chemoattractant field $C$ and fluid velocity $\mathbf{u}$ for chemotactic swimmers in the ``hydrodynamic" regime for non-tumbling swimmers $\lambda_0=0$ (top) and for purely tumbling swimmers with $\lambda_0-0.025$ for $\nu=0, 0.1$.
}
\label{fig:Tumbling}
\end{figure*}

Fig. \ref{fig:Tumbling} shows simulations of the dynamics of an initially isotropic pusher suspension with basal tumbling  with varying hydrodynamic resistance parameters $\nu=0, 0.1$. We do not show the dynamics for $\nu=0$ because the perturbations decay to zero, as expected from the stability analysis.  For comparison, we also show the dynamics in the case of the non-tumbling Stokesian swimmers.

\begin{figure*}[htpb]
\centering
\includegraphics[width=1.6\columnwidth]{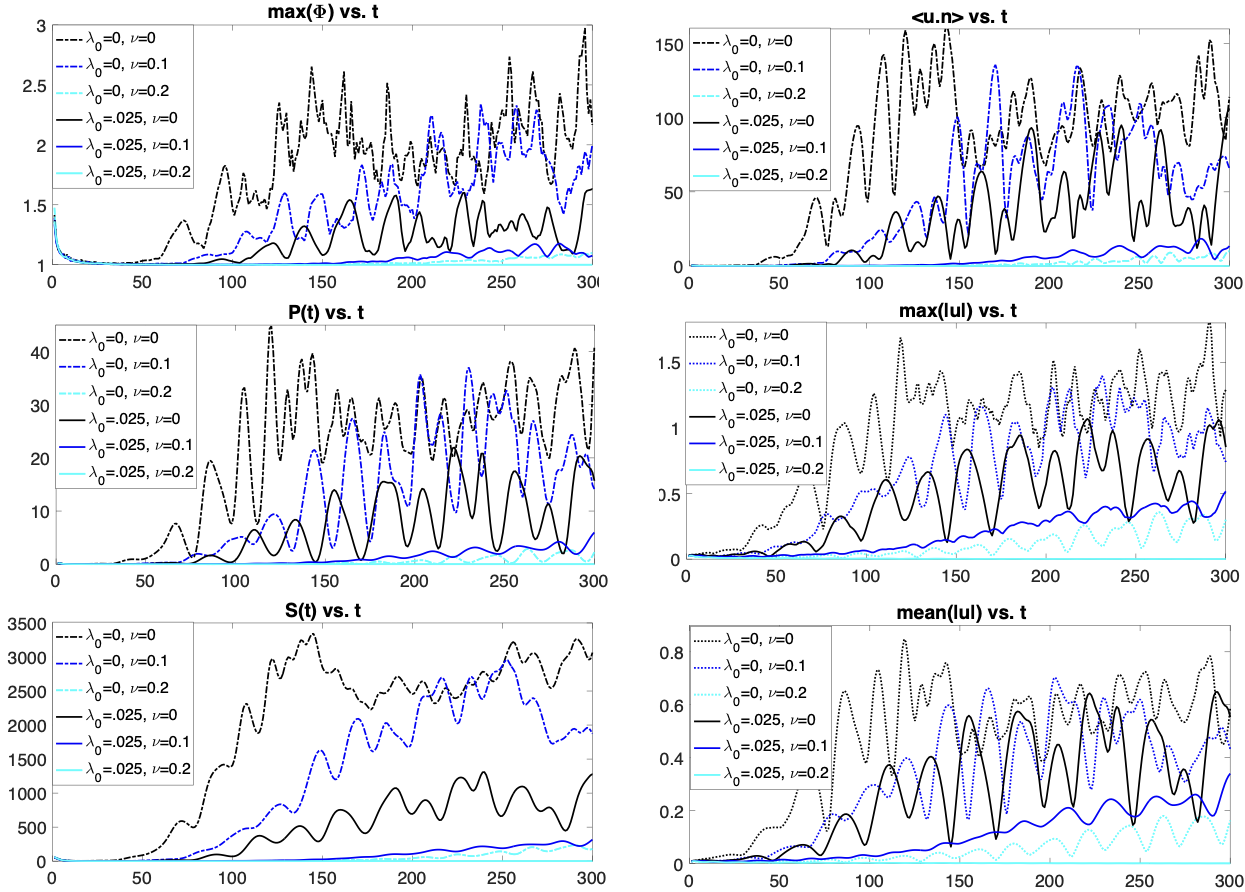}
\vspace{-0.1in}
\caption{Comparisons over a long time of $max(\mathbf{\Phi})$, $max(|\mathbf{u}|)$, $mean(|\mathbf{u}|)$, $<\mathbf{u} \cdot \mathbf{n}>$, $\mathcal{S}(t)$, and $P(t)$ for purely tumbling swimmers with $\lambda_0=0.025$ and $\nu=0, 0.1$. Results for non-tumbling swimmers with $\lambda_0=0$ are included for comparison.
}
\label{fig:AnalysisTumbling}
\end{figure*}

For low resistance $\nu=0.1$, we observe the emergence of bacteria concentration bands that exhibit bending, folding, and intricate patterns, accompanied by a fluid flow that promotes mixing--a signature of the emerging dynamics due to the hydrodynamic instability for pusher suspensions in Stokes flow \cite{Saintillan08a, Saintillan08b} that persists when including low tumbling ($\lambda_0=0.025, \nu=0)$ and now low resistance ($\nu =0.1$). However, for $\nu=0.1$ we observe a visible delay in the onset of the instability compared to the homogenous case with $\nu = 0$. This delay suggests that higher resistance hinders the development of the instability, resulting in a slower evolution of the concentration bands. 

For higher resistance, $\nu = 0.2$ (not shown), there's a significant suppression of the instability, with the initial perturbations decaying to zero and resulting into a stabilized and uniform dynamics. This occurs because for $\nu =0.2, \lambda_0=0.025$ we have $Re(\sigma_H) \approx 0$ (see the last panel of Fig. \ref{fig:ReImSigma_RT}), and the parameter set is on the margins of the ``hydrodynamical collective swimming" state in \ref{fig:Phase_slices1}. This suggests that strong hydrodynamic resistance can effectively suppress the emergence of concentration bands and the hydrodynamic collective swimming state.  Higher resistance values result in  the same, as expected for parameter values falling under the uniform dynamics region in the phase diagrams of Figs. \ref{fig:PhaseDiagram_3D}-\ref{fig:Phase_slices1}.

We can quantify the impact of Brinkman resistance by measuring over time certain metrics, e.g. the maximum swimmer concentration $\max(\Phi)$, the maximum and mean values of the fluid velocity $\max(|\mathbf{u}|)$, $\text{mean}(|\mathbf{u}|)$, the propensity of the swimmers to align with the fluid flow  $<\mathbf{u} \cdot \mathbf{n}>$ (where $\mathbf{n} =<\mathbf{p}>/\Phi$) as pushers typically do so in Stokes case \cite{Saintillan08b}. We also trace the time-evolution of the configurational entropy $\mathcal{S}(t) =  \int \int  (\Psi /\Psi_0) \ln(\Psi /\Psi_0) d\mathbf{p} d\mathbf{x}$ and the global input power $P(t)=-\int  \mathbf{E} : \Sigma^p d\mathbf{x}$, as defined in prior studies \cite{Almoteri25a, Lushi18, Saintillan13, Saintillan08b}. From the analysis of $\mathcal{S}(t)$ in \cite{Almoteri25a}, fluctuations are expected to increase in pusher swimmer suspensions, who also generate a positive input power, and are balanced by the resistance, tumbling and diffusion. 

Fig. \ref{fig:AnalysisTumbling} shows the time-evolution of these quantities for different resistance values. For comparison, we also include the non-tumbling $\lambda_0=0$ case, whose parameters are shown as blue dots in the diagrams Figs. \ref{fig:PhaseDiagram_3D}-\ref{fig:Phase_slices1}.

For pusher suspensions these quantities initially increase and eventually saturate because the active input power is balanced by the diffusive processes, as in Stokes case \cite{Saintillan08a, Saintillan08b}. Without tumbling, we observe a decrease in the saturated magnitude of all these quantities for increasing resistance $\nu$. Upon introducing tumbling $\lambda_0$ into the system, we see a further decrease in the magnitudes of all these quantities compared to the non-tumbling case. Basic tumbling per se, as anticipated from the linear analysis and Fig. \ref{fig:ReImSigma_RT}, acts as a stabilizer on the system. Resistance dampens the hydrodynamic instability and  the emergence of the collective swimming state, and can fully suppress it if sufficiently large, e.g. as with $\lambda_0=0.025, \nu=0.2$.

\vspace{-0.1in}
\subsection{Resisting dynamic aggregation}
\vspace{-0.1in}

We next check the effect of resistance on the dynamic aggregation state--where {\em both} the chemotactic instability and the hydrodynamic instability for pushers are present. We examine the dynamics of auto-chemotactic pusher suspensions with parameters $\lambda_0=0.025$, $\chi=50$, $\beta_1=0.25$, $\beta_2=0.25$, and $D_c=0.05$ for which there is both a hydrodynamic instability and an auto-chemotactic instability for the chosen domain size $L=25$, as shown in Figs. Fig. \ref{fig:ReImSigma_RT}, \ref{fig:Sigma_C}. This parameter set is denoted by red dots in the phase space diagrams, Fig. \ref{fig:PhaseDiagram_3D}, and is carefully chosen to reside within the dynamic aggregation region for $\nu=0, 0.1$, barely for $\nu=0.2$, but not for $\nu=0.3$. 

\begin{figure*}[htpb]
\centering
\includegraphics[width=1.8\columnwidth]{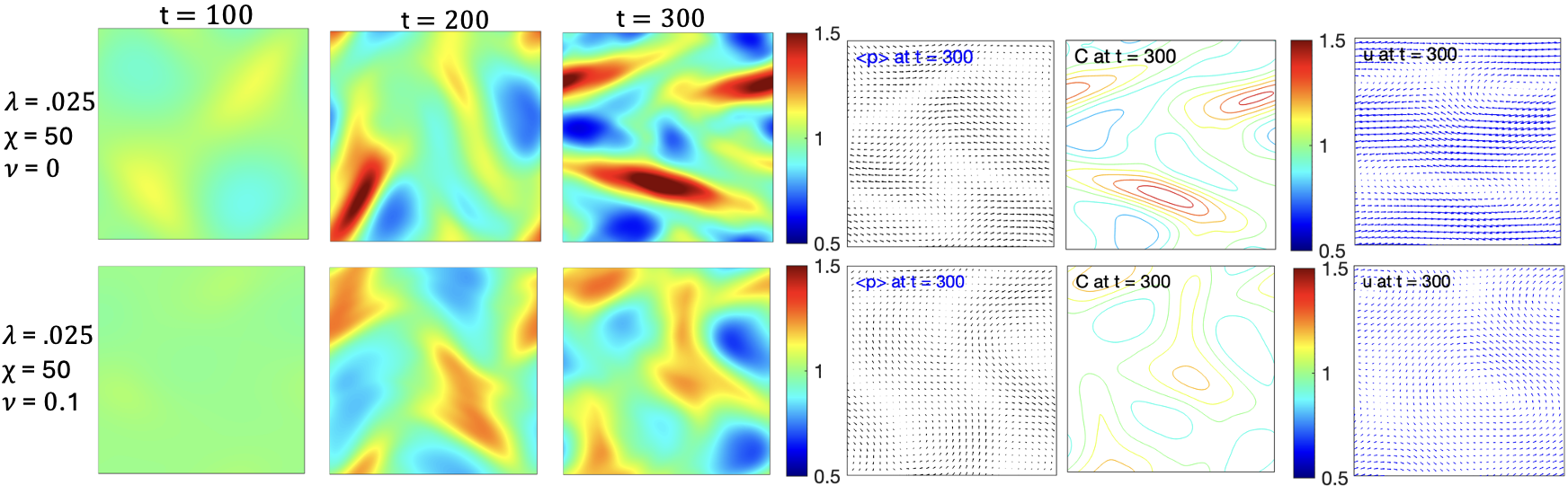}
\vspace{-0.1in}
\caption{The {\bf pusher} swimmer concentration $\Phi$ at times $t=100,200,300$, the swimmer director $<\mathbf{p}>$, the chemoattractant field $C$ and fluid velocity $\mathbf{u}$ for chemotactic swimmers in the ``dynamic aggregation" regime for $\nu=0, 0.1$. 
}
\label{fig:Chemotaxis1}
\end{figure*}

\begin{figure*}[htpb]
\centering
\includegraphics[width=1.6\columnwidth]{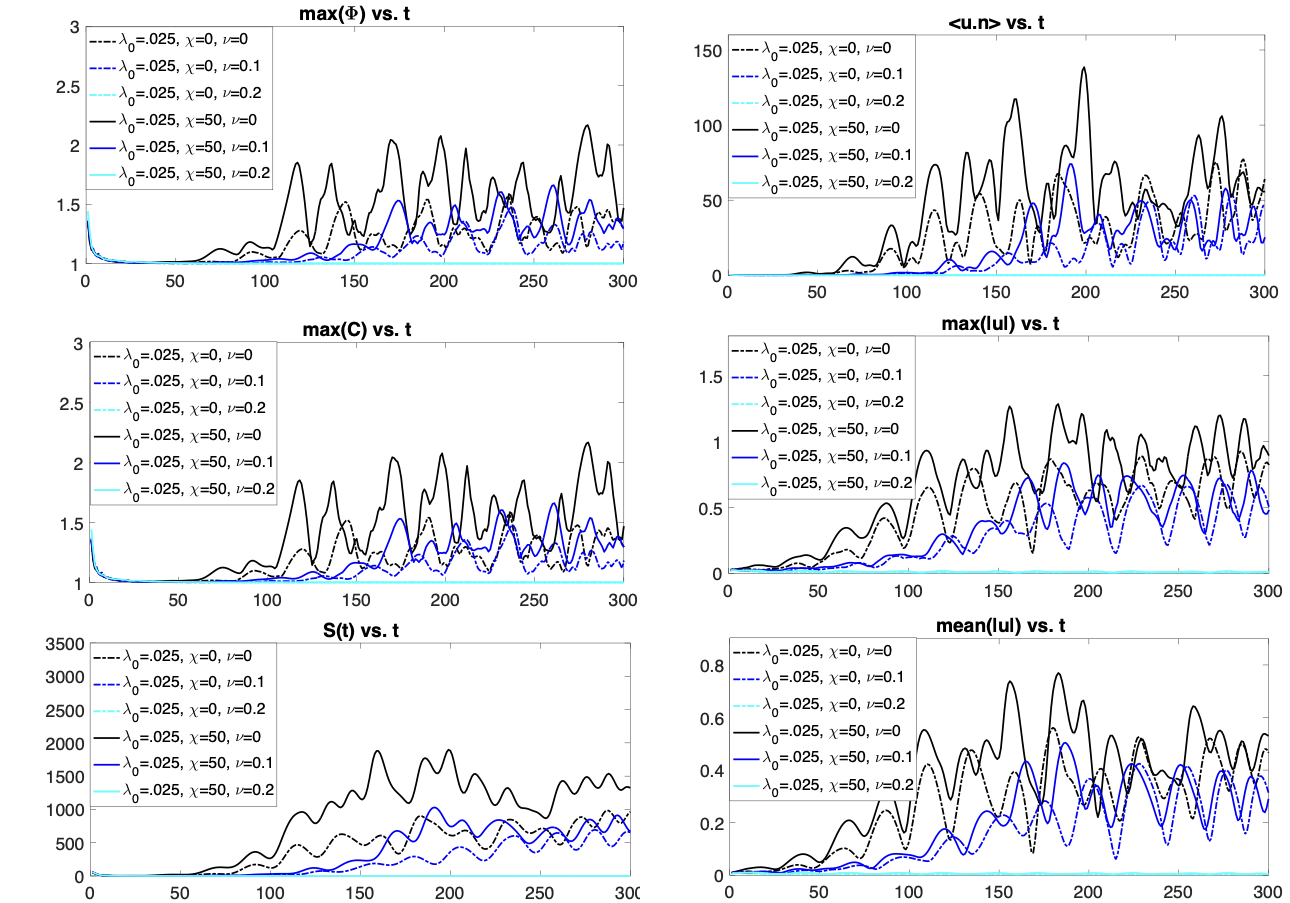}
\vspace{-0.1in}
\caption{Comparisons of $max(\mathbf{\Phi})$, $max(|\mathbf{u}|)$, $mean(|\mathbf{u}|)$, $<\mathbf{u} \cdot \mathbf{n}>$, $\mathcal{S}(t)$, and $P(t)$ for chemotactic swimmers in the ``dynamic aggregation" regime for $\nu=0, 0.1, 0.2$.  Results for non-chemotactic swimmers $\lambda_0=0.025, \chi=0$ are included for comparison.
 }
\label{fig:AnalysisChemotaxis1}
\end{figure*}

Fig. \ref{fig:Chemotaxis1} presents snapshots of the dynamics that allow us to observe firstly the predicted dynamic aggregation state and changes in the suspension dynamics as we increase the hydrodynamic resistance parameter $\nu$.

First, we describe the dynamics pertaining to the dynamic aggregation state in homogenous flows $\nu=0$, as also studied by \cite{Lushi18}.  The micro-swimmers produce chemo-attractant as well as aggregate towards it. A strong mixing flow emerges, and it advects both the swimmers and the chemo-attractant, resulting in dynamic aggregation of swimmers occurring due to their local auto-chemotactic tendency. This effect is seen from the sharper and narrower concentration bands in the auto-chemotactic suspension in Fig. \ref{fig:Chemotaxis1} compared to the non-chemotactic or just tumbling cases presented earlier in Fig. \ref{fig:Tumbling} for the same $\nu=0$.

In the rest of the figures, we can clearly see the impact of increasing resistance on the swimmer suspension dynamics. One noticeable effect is the visible delay in the onset of the instability as we increase $\nu$. This delay indicates that the higher resistance hinders the development of the combined instability, resulting in a slower evolution of the concentration bands. 

It is interesting to note that for $\nu = 0.2$ (not shown), despite the parameter set falling within the dynamic aggregation region, we observe uniform dynamics instead of the expected aggregation behavior. This occurrence highlights the concept of marginal stability due to the proximity of the parameter set location to both the region boundary surfaces $\lambda_0=\lambda_0(\nu=0.2)$ and $1/\lambda_0 h(\nu)= \chi \beta_2/\beta_1$. Moreover, the phase diagram does not include the effects of the translational and the rotational diffusions that are present in the simulations. This result also reminds us that linear theory alone cannot predict the nonlinear system dynamics. As expected, for $\nu >0.2$ we obtain the uniform dynamics state.

As before, we can quantify these observed effects by comparing quantities like the maximum concentration, maximum and mean fluid velocity, entropy and global input power, shown in Fig.\ref{fig:AnalysisChemotaxis1}. For comparison, we also include the quantities for a non-chemotactic suspension with the same basic tumbling rate.

Comparing the chemotactic suspension to the purely tumbling one of the same $\nu$, we observe an overall higher magnitude of all the measured quantities. Notably, at the onset of the instability, the quantities' highs and lows seem to occur at the same time, supporting the idea that here auto-chemotaxis reinforces the dynamics induced by the hydrodynamic collective swimming.

Increasing the resistance $\nu$ we notice an overall decrease in all the quantities, even though the chemotactic instability should still be somewhat present and barely affected according to the linear stability. This can be attributed to the resistance's damping effect to the overall swimmer motion, which in turn affects aggregation due to the nonlinear coupling of the terms. 

Lastly, even though linear theory gave separate hydrodynamic and chemotactic instabilities and predicted resistance to only significantly affect the former, simulations where they are fully and nonlinearly couple clearly show that it affects both processes in the dynamic aggregation regime.

\vspace{-0.1in}
\subsection{Resisting auto-chemotactic aggregation}
\vspace{-0.1in}

Here we focus on investigating the effects of resistance on the auto-chemotactic aggregation regime of Fig. \ref{fig:PhaseDiagram_3D}, where according to the linear theory only the chemotactic instability is present. We select parameters $\lambda_0=6, \chi=2, \beta_1=0.1=\beta_2$, and $D_c=0.4$. These parameter values are of particular interest as they approximate conditions from experimental studies involving chemotactic E. Coli bacteria \cite{Lushi18, Saragosti11, Saragosti10}. 

The selected parameter values yield a chemotactic instability according to the linear stability analysis as shown in the plot of $\sigma_C$ in Fig. \ref{fig:Sigma_C}, hence from it alone we expect aggregation and clustering of the swimmers. Note that $\lambda_0=6.0$ is considerably larger than the threshold value of $\lambda_0=0.2$ needed to  fully suppress the hydrodynamic collective swimming according to the linear stability analysis. This parameter set is too far to the right and not shown in the phase diagram  Fig. \ref{fig:PhaseDiagram_3D}. 

\begin{figure*}[htpb]
\centering
\includegraphics[width=1.8\columnwidth]{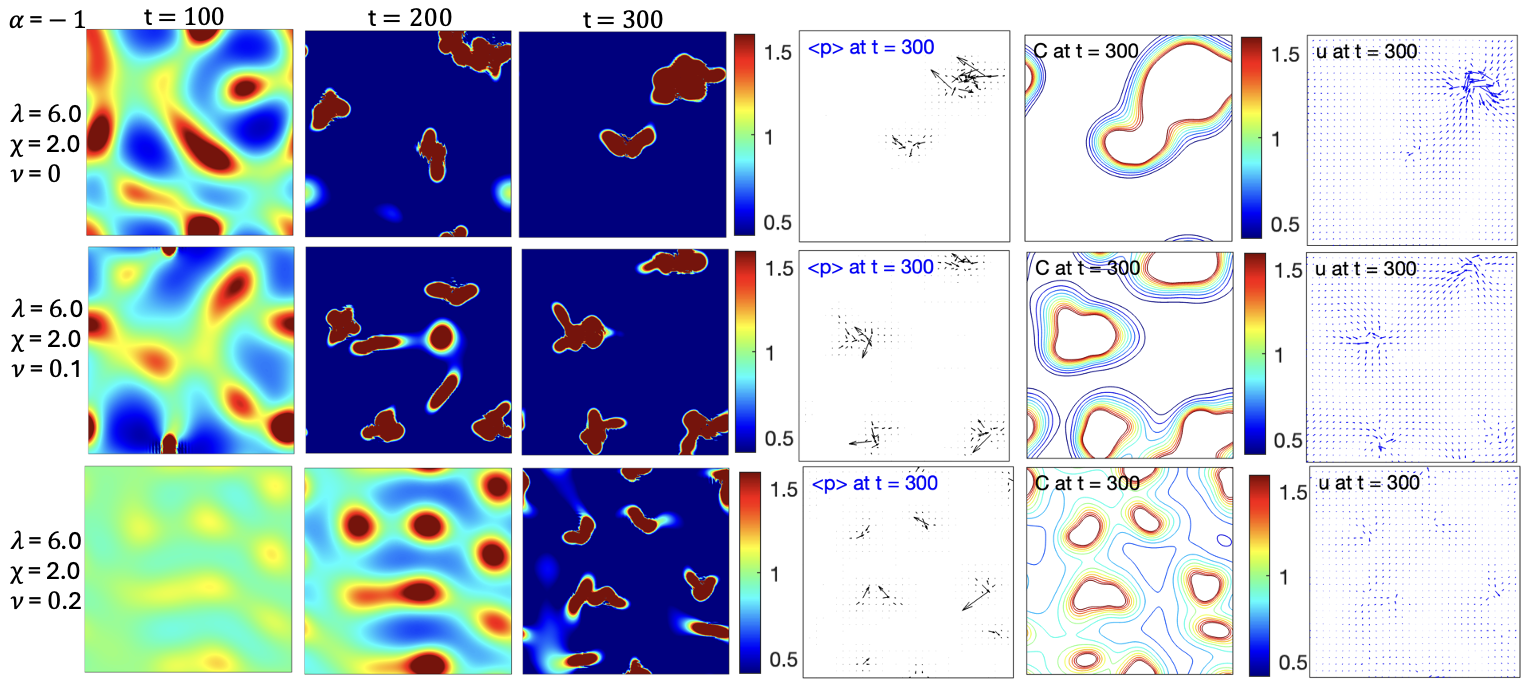}
\vspace{-0.1in}
\caption{The {\bf pusher} swimmer concentration $\Phi$ at times $t=100,200,300$, the swimmer director $<\mathbf{p}>$, the chemoattractant field $C$ and fluid velocity $\mathbf{u}$ for chemotactic swimmers in the `` auto-chemotactic aggregation" regime for $\nu=0, 0.1, 0.2$. 
}
\label{fig:Chemotaxis2}
\end{figure*}

In Fig. \ref{fig:Chemotaxis2}, we present snapshots of the average swimmer director, concentration field, generated fluid flows (not to scale), and chemical concentration for various resistance parameters $\nu$.

\begin{figure*}[htpb]
\centering
\includegraphics[width=1.6\columnwidth]{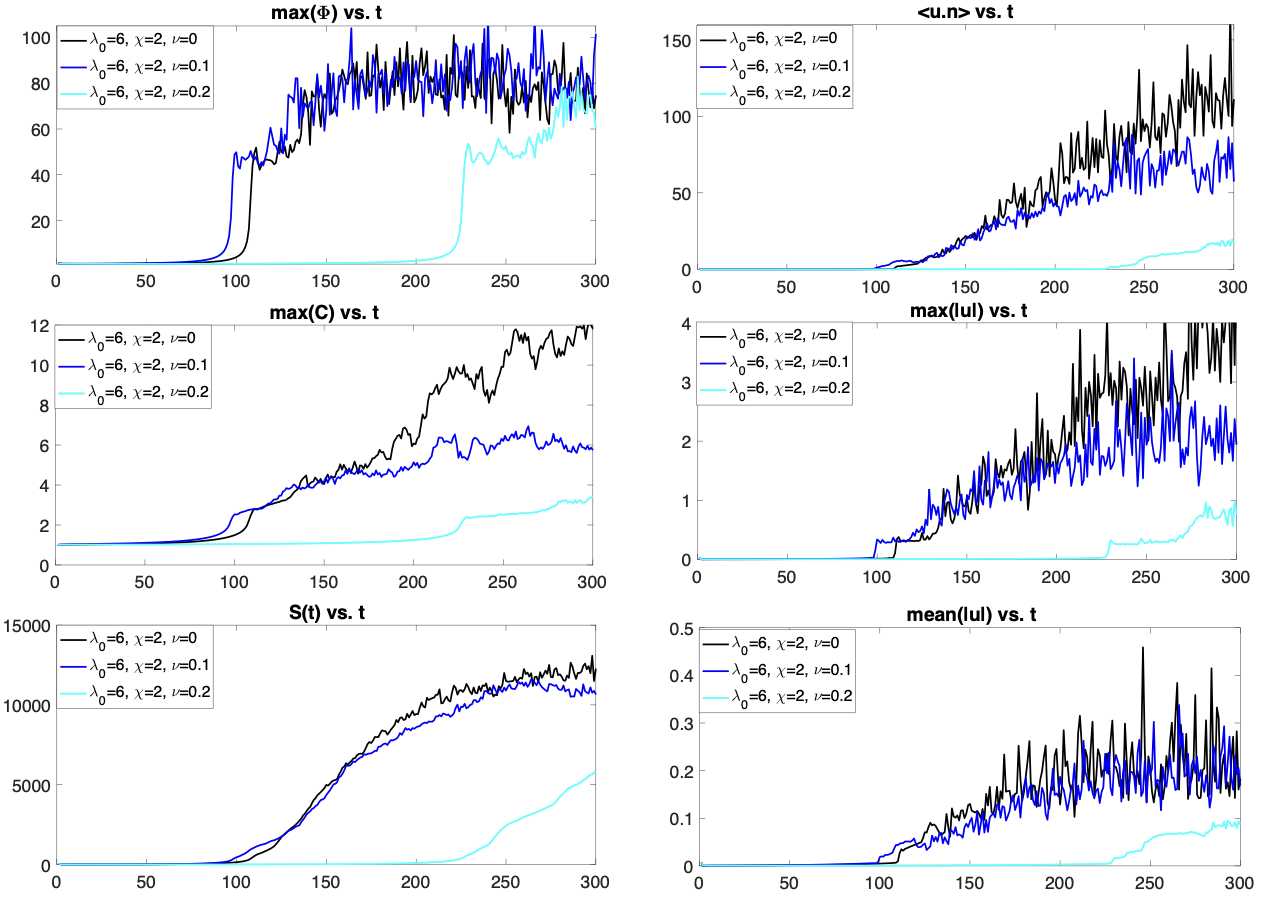}
\vspace{-0.1in}
\caption{Comparisons of $max(\mathbf{\Phi})$, $max(|\mathbf{u}|)$, $mean(|\mathbf{u}|)$, $<\mathbf{u} \cdot \mathbf{n}>$, $\mathcal{S}(t)$, and $P(t)$ for chemotactic swimmers in the ``chemotactic aggregation" regime for $\nu=0, 0.1, 0.2$.
}
\label{fig:AnalysisChemotaxis2}
\end{figure*}

First, let us explain the aggregation dynamics that emerges in the absence of resistance, Fig. \ref{fig:Chemotaxis2} top. The perturbations in this pusher suspension initially merge into quasi-elliptical clusters, which then start to move and become dynamic. These clusters continuously change shape, move around the domain, and may merge with other motile clusters \cite{Lushi18}. Despite this parameter set being located far from the hydrodynamic instability region in the parameter phase space, the fluid flows in these aggregates are nontrivial due to the high swimmer concentration there and the nonlinear coupling. In fact, it is these local fluid flows that contribute to the deformation of the clusters as well as their motility \cite{Lushi18}. 

Examining snapshots of the dynamics for $\nu>0$ in Fig. \ref{fig:Chemotaxis2}, we observe that resistance appears to inhibit auto-chemotactic aggregation into larger clusters. As $\nu$ increases, we observe more and smaller-sized motile aggregates. It appears that, as resistance lowers individual swimmer motility, it hinders their ability to move towards clusters and join other swimmers in those aggregates, resulting overall in more clusters of smaller size.

As before, we quantify these observed effects of the Brinkman resistance in the formation and sustenance of these motile pusher aggregates in Fig.\ref{fig:AnalysisChemotaxis2}. They confirm our observations in the dynamics snapshots. Moreover, we note that there is a delay on the onset of the non-trivial dynamics for $\nu=0.1$ compared to $\nu=0$, and the delay for the onset of the dynamics for $\nu=0.2$ is much more pronounced.  

It is important to note that in the linear theory analysis, resistance was present in both the hydrodynamic dispersion relation and the chemotactic dispersion relation, even though the effect in the latter appeared to be minimal and almost negligible. From prior work on non-chemotactic suspensions, the effect of the resistance in pusher suspensions was primarily through the bulk-fluid effects, i.e. the inclusion of the $-\nu^2 \mathbf{u}$ term in the fluid equations instead of the slowdown in the individual swimmer $U_B$, which is  also how it influences the chemotactic instability here (through $\mathcal{D}_t C$). This implied that resistance would primarily influence the hydrodynamic collective swimming state and the dynamical aggregation state. However, in the full system where the resistance, hydrodynamics and chemotaxis becomes intricately coupled in non-linear terms, we see that resistance also significantly impacts the chemotactic aggregation. Resistance hinders the formation of clusters by impeding the individual swimmer motion and hydrodynamic interaction with others, and ultimately affects the swimmers' ability to aggregate.

Last, but not least, we note that we did elaborate here on the effects of resistance in the chemotactic dynamics of {\em puller} suspensions. From prior work on Stokesian suspensions we know that the collectively-generated fluid flows, though non-trivial, can act as separators in chemotactic clusters to keep them from further merging \cite{Lushi12, Lushi18}. This will be investigated in more detail in subsequent work.

\vspace{-0.1in}
\section{Parameter Values}
\vspace{-0.1in}

In \cite{Almoteri25a} we presented some values of the parameters gleaned from various experimental studies, namely \cite{Vanni00, Cortez10, Drescher11, Dombrowski04, Tuval05, Aranson07, Sokolov07, Sokolov09, Cisneros11, Sokolov12}. We repeat them in the table below and add the chemotactic ones from \cite{Saragosti11, Saragosti10}. The reader can infer the non-dimensional parameter values by following the non-dimensionalization presented in Section IIB and our prior work in \cite{Almoteri25a} .

{ \footnotesize
\begin{table*}[ht]
\centering
\begin{tabular}{l | c | c | c | c | c | c}
 Parameter & $l$  &  $U_0$ & $n$  & $\ell_c$ & $|\sigma_0|$ & $\mu$ \\
Exp. value & $4$-$5 \mu m$  & $20\mu m s^{-1}$  &  $0.5$-$2*10^{10} cells cm^{-3}$ & $3$-$12.5 \mu m$ & $1.6$-$2.1 pN \mu m$ & $10^{-3} pNs \mu m^{-2}$ \\ 
 \hline
 Parameter &  $D_t$  & $D_r$  & $K_D$  & $\beta_2$ & $\beta_1$ & $D_c$\\
Exp. value & $200 \mu m^2 s^{-1}$ & $0.1 rad^2 s^{-1}$  & $125$-$2007 \mu m^2$ & $4*10^5 cell^{-1} s^{-1}$ & $4*10^{-3} s^{-1}$ & $8*10^{-6} cm^2 s^{-1}$
\end{tabular} 
\end{table*}
}

This continuum model is built upon considering dilute swimmer and dilute obstacle suspensions, and it does not include direct swimmer-swimmer and swimmer-obstacle interactions which would have a significant impact on the dynamics. Thus we add the caveat that, as with most such theoretical models that cannot include all the complexities of experimental environments, the results here should be viewed qualitatively and not quantitatively,  but hopefully still motivate new studies.

\vspace{-0.1in}
\section{Summary and Discussion}
\vspace{-0.1in}

Through analysis and simulations of a continuum model, we studied the dynamics of a dilute suspension of autochemotactic micro-swimmers in Brinkman fluids. The model consists of a conservation equation for the swimmer positions and orientations that includes a chemotactic run-and-tumble response as well as swimmer advection and rotation by the fluid. This is coupled to the chemo-attractant dynamics, which includes production by the swimmers and advection by the fluid flows. The fluid flow equations include an active particle stress to account for the swimmers' motion in it and a linear resistance or friction term to account for the medium inhomogeneity or ``porosity''.  In addition, the model includes the slow-down in the individual swimmer speed that results from the medium resistance.

We analyzed the stability of perturbations from the uniform isotropic state of swimmers suspensions with a quasi-static chemo-attractant field. The linear analysis reveals two separate dispersion relations or implicit equations for the growth rate of the perturbations: one resulting from the hydrodynamic interactions between swimmers and the other resulting from the swimmer autochemotaxis, with the resistance parameter appearing in both. Asymptotic analysis for small wavenumbers revealed two branches for the hydrodynamics-related instability, which could be positive for pusher swimmers, and one branch for the chemotaxis-related instability. Resistance appears in the growth rate expressions for both instabilities and has a stabilizing effect. 

The linear stability analysis helps us identify parameter ranges for which the hydrodynamic and autochemotactic instabilities are to be expected for elongated pusher swimmers, and we gather this information in a phase diagram. Accordingly, we identify four dynamics states; namely {\em hydrodynamic collective swimming} due to the expected hydrodynamic instability,  {\em chemotactic aggregation} due to the expected autochemotactic instability, {\em dynamic aggregation} due to the presence of both instabilities, and {\em uniform} state due to trivial dynamics expected by the absence of both instabilities. 

Numerical simulations of the full nonlinear equations for parameters in the three non-trivial dynamics states for varying resistance values confirm the linear analysis results that overall resistance has a stabilizing effect in pusher swimmer suspensions, and if it is sufficiently large, it can completely suppress these instabilities. In the heterogeneous Brinkman flow cases we see a qualitatively similar dynamics to the Stokesian case, except the onset of either instability is typically delayed and the concentration fluctuations weaken with increased resistance values. The resistance hinders the onset as well as the development of collective swimming in pusher swimmer suspensions as it dampens the hydrodynamic instability and can completely suppress it altogether if larger than a value close to the critical one predicted by linear theory. This is the case also in the dynamical aggregation regime where autochemotaxis typically reinforces the swimmer concentration bands: resistance hinders the autochemotactic collective swimming. In the chemotactic aggregation state, even far from the regimes where linear analysis predicted the hydrodynamic instability to occur, we still see the emergence of non-trivial fluid flows in clusters with high swimmer concentration, and as such resistance can also have an effect. In these cases, resistance hinders the ability of the swimmers to move towards other swimmers and as such impedes aggregation, with the result manifesting as more swimmers clusters of smaller size.

Though our system has not been realized in any experiments, we present experimental parameter values in the hope that they will stimulate discussion and inspire new work. Promising experimental setups include bead-like gels \cite{Bhattacharjee19a, Bhattacharjee19, Bhattacharjee22}, quasi-2D disordered environments \cite{Makarchuk19, Dehkharghani23}, or colloid suspensions \cite{Kamdar22}, even though they have micro-structures of a scale larger than can be approximated with our Brinkman model.  

Our model is limited to dilute chemotactic swimmer suspensions, but it is possible to add an aligning effect to account for higher swimmer densities \cite{Ezhilan13}. The system can be adapted to study swimmer dynamics in environments with space-dependent friction or patterned surfaces, as in \cite{Volpe11,Stoop19, Thijssen21}, and help to devise novel ways to control the flow and transport of active suspensions. 

\section{Acknowledgements}
The authors gratefully acknowledge support from the Simons Foundation (EL) and fellowships from the Kingdom of Saudi Arabia (YA), and thank T. Bhattacharjee, S. Datta, N. Desai, G. Elfring, F. Guzman-Lastra, E. Lauga, A. Morozov, T. Pedley, D. Pushkin, R. Soto and S. Spagnolie for helpful discussions. 
EL thanks the Newton Institute for Mathematical Sciences, Cambridge, for inclusion and hospitality during the program  ``Anti-diffusive dynamics: from sub-cellular to astrophysical scales'' supported by EPSRC grant no. EP/R014604/1.

\bibliography{references}

\end{document}